**Extending the capillary wave model to include the effect of bending rigidity: X-ray reflection and diffuse scattering.**

Chen Shen (沈辰),[1,*] 0000-0002-7855-1764
Honghu Zhang (张洪湖),[2] 0000-0003-1784-7825
Beate Klösgen,[3,1] 0000-0002-2334-0388
Benjamin M. Ocko[2,†] 0000-0003-2596-1206

[1]Deutsches Elektronen-Synchrotron DESY, Notkestr. 85, 22607 Hamburg, Germany
[2]National Synchrotron Light Source II, Brookhaven National Laboratory, Upton, NY 11973, USA
[3]Department of Physics, Chemistry and Pharmacy, University of Southern Denmark, Campusvej 55, 5230 Odense, Denmark

*Contact author: chen.shen@desy.de
†Contact author: ocko@bnl.gov

**ABSTRACT**. The surface roughness of a thin film at a liquid interface exhibits contributions of thermally excited fluctuations. This thermal roughness depends on temperature (T), surface tension ($\gamma$) and elastic material properties, specifically the bending modulus ($\kappa$) of the film. A non-zero $\kappa$ suppresses the thermal roughness at small length scales compared to an interface with zero $\kappa$, as expressed by the power spectral density (PSD) of the thermal roughness. The description of the X-ray scattering of the standard Capillary Wave Model (CWM), that is valid for zero $\kappa$, is extended to include the effect of $\kappa$. The extended CWM (eCWM) provides a single analytical form for both the specular X-ray reflectivity (XRR) and the diffuse scattering around the specular reflection, and recovers the expression of the CWM at its zero $\kappa$ limit. This new theoretical approach enables the use of single-shot grazing incidence X-ray off-specular scattering (GIXOS) measurements for characterizing the structure of thin films on a liquid surface. The eCWM analysis approach decouples the thermal roughness factor from the surface scattering signal, providing direct access to the intrinsic surface-normal structure of the film and its bending modulus. Moreover, the eCWM facilitates the calculation of reflectivity at any desired resolution (pseudo XRR approach). The transformation into pseudo XRR provides the benefit of using widely available XRR software to perform GIXOS analysis. The extended range of the vertical scattering vector ($Q_z$) available with the GIXOS-pseudo XRR approach allows for a higher spatial resolution than with conventional XRR. Experimental results are presented for various lipid systems, showing strong agreement between conventional specular XRR and pseudo XRR methods. This agreement validates the proposed approach and highlights its utility for analyzing soft, thin films.

## I. INTRODUCTION

Soft matter thin films are abundant in nature, and they may serve as important engineering materials with well-defined functional properties. Typical examples are biomimetic and biological membranes [1,2], materials for drug delivery [3-5], surfactants for ion floatation [6], and 2-dimensional conjugated polymer networks for organic electronic applications [7]. Exploring the structures of such films and their phase behavior at nanoscopic length scales is pivotal for understanding their functional properties, and an important basis for rational design of novel materials. The physical and chemical conditions of the surface of a liquid substrate renders it advantageous over the alternative solid substrate for studies of soft matter thin films for several reasons [1,8]. Firstly, there is no underlying substrate pinning effect that reduces the lateral mobility in the film. The high mobility within the layer and its low energy barriers are relevant to assess fundamental processes like interfacial (un-)binding [2,9], raft formation, and even fusion of biological membranes [5,10]. Further, the underlying aqueous chemistry can be controlled and adjusted to precisely tune substrate-film interactions, a feature almost impossible on solid supported (multi-)layers. A typical example is the use of a Langmuir trough where thermodynamic properties like surface pressure and subphase compositions, e.g. ionic strength, are varied during studies of the biomimetic monolayers [1].

Surface scattering techniques using X-rays and neutrons are well established and widely used methods to acquire the structure of liquid interfaces [8,11-21]. The time-averaged structure along surface normal is commonly obtained through conventional X-ray (XRR) and neutron reflectivity (NR) methods: analyzing the measured angular dependence of the reflected intensity [11]. With these conventional methods, the spatial resolution of the obtained structure is affected by thermally excited capillary wave fluctuations that give rise to an interfacial thermal roughness. This roughness broadens the surface normal density profile which often makes it difficult to resolve the depth distribution of chemical moieties. As shown previously [12,16,18-20,22-24], improved spatial resolution is achieved by analyzing the diffuse scattering profile around the specular peak. In the case of liquid surfaces with zero bending rigidity, the thermal roughness contribution to the specular reflection and the surrounding diffuse scattering is accurately described by the Capillary Wave Model (CWM) [18]. For such systems, the intrinsic structure factor along the surface normal is obtained directly by separating the thermal roughness contribution from the scattering signal and this improves the spatial resolution of the reflectivity analysis [12,18,24].

In the more realistic case of a film with non-zero rigidity at a liquid interface, the CWM fails to properly account for the thermal roughness and hence the measured surface diffuse scattering [13-15,25,26]. This is because the stiffness of the film suppresses the amplitudes of the thermal fluctuations at the shortest wavelengths, compared to their amplitudes as predicted by the CWM [26,27]. The observed deviation of the diffuse scattering from the CWM prediction suggests that a non-zero bending modulus $\kappa$ has an impact not only for amphiphile monolayers on water [13-15,25,26,28] but also for surface frozen alloy phases [29]. For such cases, the CWM needs to be extended to correctly evaluate the thermal roughness and to obtain the intrinsic structure factor.

*Contact author: chen.shen@desy.de

†Contact author: ocko@bnl.gov

To date, existing theories that have incorporated bending rigidity all use approximations that are only valid for a small out-of-plane scattering vector $Q_z$ [26,29], or for a small in-plane scattering vector $Q_{xy}$ [13,14,28,30] (see FIG 1 for the scattering vector definitions). The approximation for small values of $Q_z$ cannot be used for the high spatial resolution structural analysis that involves data at large $Q_z$ [24,31]. In addition, the use of the small $Q_{xy}$ approximation also limits the applicable range in $Q_z$. This is because the diffuse scattering at small $Q_{xy}$ is often measured by a so-called "rocking scan" around the specular reflection ($\beta = \alpha$), a simultaneous scan of both the incident angle $\alpha$ and the vertical scattering angle $\beta$ in the plane of incidence ($2\theta = 0$, see FIG 1) [13,28,30], while maintaining their sum ($\alpha + \beta$) constant. Under this geometry, a large value of $\alpha$ is required to reach a large $Q_z$ ($Q_z = \frac{4\pi}{\lambda}\sin\alpha$ at specular reflection, $\lambda$ is the incident wavelength). But at such large $\alpha$, the bulk scattering is often orders of magnitude stronger than the surface scattering signal, making it difficult to measure the latter [24,31]. Hence a rocking scan is far from ideal for measuring surface diffuse scattering at large $Q_z$ [24,32].

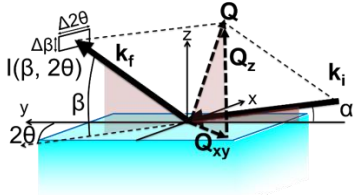

FIG 1. Schematics of the X-ray scattering geometry from a liquid surface (blue part). The sketch depicts the incident angle $\alpha$, the vertical and the horizontal scattering angle $\beta$ and $2\theta$, the wavevectors $k_i$ and $k_f$ of the incident and the scattered beam, and the in-plane and out-of-plane components $Q_{xy}$ and $Q_z$ of the scattering vector $Q$, respectively. A detector is positioned at the angle ($\beta$, $2\theta$). Scattered photons are integrated within a rectangular region, of which the size corresponds to angular openings of $\Delta\beta \times \Delta2\theta$. The figure is reproduced with the permission from [31] under CC-BY 4.0 license.

Obtaining the intrinsic surface normal structure factor at large $Q_z$ requires (1) an experimental method to measure the diffuse and/or the specular scattering to a large $Q_z$ range and (2) a functional expression for the thermal roughness contribution to the scattering signal that is applicable to this $Q_z$ range. Using grazing incidence X-ray off-specular scattering (GIXOS) under total external reflection [21], we have previously shown that the background subtracted diffuse scattering around the specular reflection can be accurately measured to large $Q_z$ (beyond 1 Å$^{-1}$) [31,33] from a liquid surface. Further, we showed from GIXOS results that the thermal roughness follows the CWM prediction for systems with zero or very small bending rigidity; in turn this allows one to obtain the reflectivity profile (pseudo reflectivity) and the intrinsic structure factor along the surface normal [31]. For films with non-zero bending moduli a similar data analysis cannot be carried out unless the functional form of the thermal roughness factor is modified with appropriate approximations. The required **Q** range is $0.01 < Q_{xy} < 0.15$ Å$^{-1}$ and $Q_z > 1$ Å$^{-1}$ and has not been covered by the previous approximations [13-15].

In this paper we propose an extension to the standard CWM that includes the effect of the bending rigidity of a film at a liquid interface. This extended Capillary Wave Model (eCWM) consists of a single analytical expression that describes both the specular reflection and the surrounding diffuse scattering. It is applicable to a thin film with stiffness at a liquid interface under the assumption that the height profile of the film follows that of the underlying liquid subsurface [15]. Specifically, it uses one unified form to describe the zero and the finite bending rigidity cases. At its zero rigidity limit this form recovers the known analytical form of the CWM [18]. We demonstrate that the previously proposed models are approximations of our eCWM expression in the small $Q$-regime [13,14,26]. We use the eCWM to analyze the GIXOS results for a few lipid systems on aqueous surfaces at several surface pressures in order to obtain the bending modulus values and their intrinsic surface normal structures. Finally, we calculate the pseudo reflectivity [31] from the GIXOS data using the eCWM, and we show that it coincides perfectly with conventional specular reflectivity measured from the same sample. The agreement between the two methods further validates the present approach.

## II. METHODS

In this part we provide a comprehensive theoretical derivation and discussion about the X-ray scattering from a thin film with a non-zero bending rigidity at a liquid interface using the eCWM. The results required to apply the eCWM to reflectivity analysis can be found in Eq. (18), Eq. (20) and Eq. (26) to (31). TABLE I and TABLE II respectively summarize the different kinds of roughness and the cut-offs that are used in the analysis.

### A. Power spectral density and height-height correlation

We review the power spectral density (PSD) and the height-height correlation of the eCWM [34] before deriving the expressions of the differential scattering cross section and the thermal roughness factor.

*Contact author: chen.shen@desy.de

†Contact author: ocko@bnl.gov

TABLE I. Roughness concepts that are introduced in the later sections. Detailed definitions are found in their first occurrence in the text.

| abbreviation | Meaning |
| --- | --- |
| $\sigma_{0,m/n}$ | intrinsic interfacial width (rms) between the slab m and n in a slab model; see section III. |
| $\sigma_{R,CWM} = \sqrt{\frac{k_B T}{2\pi\gamma} \ln \frac{Q_{max}}{\delta Q_{xy,R}}}$ | CWM thermal roughness ($\kappa = 0$), obtained from reflectivity [18]. Note that it is resolution ($\delta Q_{xy,R}$) dependent. |
| $\sigma_{R,eCWM} \approx \sqrt{\frac{k_B T}{2\pi\gamma} \ln \frac{0.89 Q_\kappa}{\delta Q_{xy,R}}}$ | eCWM thermal roughness that includes the effect of a non-zero bending modulus, obtained from reflectivity; see section II-D. Note its resolution dependence. |
| $\sigma_{R,m/n} = \sqrt{\sigma_{0,m/n}^2 + \sigma_{R,eCWM}^2}$ | phenomenological width of the interface between slab "m" and "n" in a slab model, obtained from reflectivity; see section IV. Note its resolution dependence through $\sigma_{R,eCWM}$. |

TABLE II. Cut-offs that occur in this work, in real space (distance) and in reciprocal space. They are defined in the section II-A.

| cut-off | real space | reciprocal space |
| --- | --- | --- |
| molecular cut-off (half capillary wave wavelength) | $a_m$ | $Q_{max} = \pi/a_m$ |
| bending rigidity cut-off | $L_\kappa = \sqrt{\kappa/\gamma}$ | $Q_\kappa = L_\kappa^{-1}$ |
| gravity - capillary wave cut-off | $L_g = \sqrt{\gamma/(\Delta\rho_m g)}$ | $Q_g = L_g^{-1}$ |

The X-ray differential scattering cross section depends on the scattering length density (SLD) $\rho_b(r_{xy}, z)$ in the illuminated volume where $r_{xy}$ is in the surface plane and $z$ is along the surface normal. Note that the SLD for X-ray scattering can be expressed as $\rho_b = r_e \cdot \rho_e$, where $\rho_e$ is the electron density and $r_e = 2.82 \times 10^{-15}$ m is the classical electron radius. When the height profile of a film follows the underlying liquid surface [‡][15], the SLD $\rho_b(r_{xy}, z)$ is related to the intrinsic SLD profile $\rho_{b,0}(z)$ and the surface height profile $h(r_{xy})$ as $\rho_b(r_{xy}, z) = \rho_{b,0}(z - h(r_{xy}))$. "Intrinsic" means that the SLD profile is for a perfectly flat, i.e. zero-roughness, underlying surface. Thermal fluctuations at a liquid surface generate an area dependent, surface thermal roughness and typical XRR and GIXOS measurements are sensitive to its time averaged profiles.

The PSD of a surface is the time and ensemble averaged square of the amplitude $\tilde{h}(Q_{xy})$ of the Fourier component of $h(r_{xy})$. For the eCWM it is [25,34]:

$$\langle \tilde{h}(Q_{xy}) \tilde{h}(-Q_{xy}) \rangle = \frac{1}{A} \cdot \frac{k_B T}{\Delta\rho_m g + \gamma Q_{xy}^2 + \kappa Q_{xy}^4} \quad (1)$$

$A$, $k_B$ and $T$ are respectively the X-ray illuminated area of the film, the Boltzmann constant and the temperature. $\Delta\rho_m$, $g$ and $\gamma$ are respectively the mass density difference across the interface, the gravitational acceleration constant and the surface tension. $\langle ... \rangle$ denotes the ensemble average. The height-height correlation $\langle h(r_{xy}) h(0) \rangle$ is computed from a 2-dimensional (2D) Fourier transform of the PSD (APPENDIX A) [34]:

$$\begin{aligned}
&\langle h(r_{xy}) h(0) \rangle \\
&= \frac{A}{(2\pi)^2} \int_0^{Q_{max}} \langle \tilde{h}(Q_{xy}) \tilde{h}(-Q_{xy}) \rangle \cdot \exp(-i Q_{xy} \cdot r_{xy}) dQ_{xy} \\
&= \frac{k_B T}{2\pi\gamma} \left[ K_0 \left( \frac{r'_{xy}}{L_g} \right) - K_0 \left( \frac{r'_{xy}}{L_\kappa} \right) \right]
\end{aligned} \quad (2)$$

$Q_{max} = \pi/a_m$ is the upper wavenumber cut-off for the shortest wavelength of the thermal roughness given by the average intermolecular distance $a_m$ and $r'_{xy} = \sqrt{r_{xy}^2 + Q_{max}^{-2}}$ [35]. $K_0(...)$ is the modified Bessel function (also called hyperbolic Bessel function) of the 2$^{nd}$ kind of the 0$^{th}$ order. In the following, two additional cut-offs are introduced. The first cut-off $L_g = \sqrt{\gamma/(\Delta\rho_m g)}$ is the large distance gravity – capillary wave cut-off below which capillary waves dominate and the modes with longer wavelength cannot be excited [35]. The second is a short distance cut-off $L_\kappa = \sqrt{\kappa/\gamma}$ involving the bending rigidity. It typically ranges between a molecular length scale up to hundreds of Ångströms (i.e. at $\kappa \sim 1000 k_B T$). In reciprocal space, the respective related wavenumber cut-offs are $Q_g = L_g^{-1}$ and $Q_\kappa = L_\kappa^{-1}$. These cut-offs are compiled in TABLE II.

Before continuing, we discuss the physical meaning of the cut-offs $L_\kappa$ and $Q_\kappa$ since they do not exist in the height-height correlation for the CWM, $\langle h(r_{xy}) h(0) \rangle_{CWM} = \frac{k_B T}{2\pi\gamma} K_0 \left( \frac{r'_{xy}}{L_g} \right)$ [35,36]. The length $L_\kappa$ represents the length scale below which the amplitudes of the capillary waves are suppressed by the stiffness at the interface in the eCWM [13,29], and this is demonstrated by the dependence of the PSD versus $Q_{xy}$ (FIG 2). At small $Q_{xy}$, much less than $Q_\kappa = L_\kappa^{-1}$, the PSD of the eCWM has the same amplitude as the one of the CWM ($\kappa = 0$, dash-dotted line) and we refer to this range as the CWM regime. For $Q_{xy}$ larger than $Q_\kappa$, the PSD of the eCWM follows a $Q_{xy}^{-4}$ dependence (black-dashed line) and is lower than the PSD of the CWM, a manifestation of the bending rigidity. Note that a large wavenumber in the reciprocal space corresponds to a shorter wavelength in the real space. This PSD behavior shows that the thermal fluctuations with wavelengths shorter than $L_\kappa$ are suppressed by the bending rigidity of the film. For the CWM the thermal roughness is only suppressed below the molecular cut-off $a_m/\pi$. [13,15,29]. Increasing the bending modulus shifts the cross-over further into to the CWM regime to lower $Q_{xy}$, and decreases the amplitude of the PSD above $Q_\kappa$ (FIG 2, black vs red solid line).

---

[‡] This is often referred as "conformal film roughness" or "fully correlated roughness between the all the interfaces".

*Contact author: chen.shen@desy.de

†Contact author: ocko@bnl.gov

The root-mean-squared (rms) height displacement $\langle h^2 \rangle$ corresponds to the surface thermal roughness averaged over a length scale larger than $L_g$ and it is obtained from the zero-distance limit ($r_{xy} = 0$) of the height-height correlation function of the eCWM:

$$\langle h^2 \rangle = \langle h(\mathbf{r}_{xy})h(\mathbf{0}) \rangle \big|_{r_{xy}=0}$$
$$\approx \frac{k_B T}{2\pi\gamma}\left[\ln\left(\frac{L_g Q_{max}}{0.89}\right) - K_0\left(\frac{1}{L_\kappa Q_{max}}\right)\right]$$
(3)

Here the approximation $\lim_{x\to 0} K_0(x) = -\ln x + \ln 2 - \gamma_E \approx -\ln(0.89x)$ is applied to the first $K_0$ term in Eq. (2) since when $r_{xy} = 0$, $r'_{xy} \ll L_g$ and $\frac{r'_{xy}}{L_g} \to 0$. $\gamma_E \approx 0.577$ is the Euler-Mascheroni constant, and the factor 0.89 originates from $\exp(\gamma_E - \ln 2) \approx 0.89$.

In the limit of zero $\kappa$, $\frac{1}{L_\kappa Q_{max}} \to \infty$, and the rms height expression of the eCWM converges to the expression of the CWM [35,37]:

$$\langle h^2 \rangle_{CWM} \approx \frac{k_B T}{2\pi\gamma}\ln\left(\frac{L_g Q_{max}}{0.89}\right) \approx \frac{k_B T}{2\pi\gamma}\ln(L_g Q_{max})$$
(4)

since $\lim_{x\to\infty} K_0(x) = 0$.

With a not so small bending rigidity such that $L_\kappa > 2Q_{max}^{-1}$, the approximation $\lim_{x\to 0} K_0(x) \approx -\ln(0.89x)$ can be applied to $K_0\left(\frac{1}{L_\kappa Q_{max}}\right)$ in Eq. (3) and the factors 0.89 in the two logarithmic terms cancel each other. The rms height displacement is simplified to:

$$\langle h^2 \rangle \approx \frac{k_B T}{2\pi\gamma}\ln\left(\frac{L_g}{L_\kappa}\right) = \frac{k_B T}{2\pi\gamma}\ln(L_g Q_\kappa)$$
(5)

The bending rigidity criterion for applying this simplified form is $\kappa > \gamma\left(\frac{2a_m}{\pi}\right)^2$, derived from $L_\kappa > 2Q_{max}^{-1}$. For typical surfactants with $a_m$ of ~ 5 Å and a tension less than 73 mN/m, the eCWM is valid when the rigidity $\kappa > 2\,k_B T$. Eq. (3) is applicable to the whole rigidity range including the range of $0 \leq \kappa \leq \gamma\left(\frac{2a_m}{\pi}\right)^2$.

A comparison of Eq. (5) (eCWM) and Eq. (4) (CWM) shows that the rms height displacement of the eCWM is smaller than that of the CWM at the same temperature and surface tension values ($Q_\kappa < Q_{max}$). This is because the finite bending rigidity limits the thermal excitation of short-wavelength height displacement. The more rigid the film, the smaller the $Q_\kappa$ and therefore the smaller $\langle h^2 \rangle$.

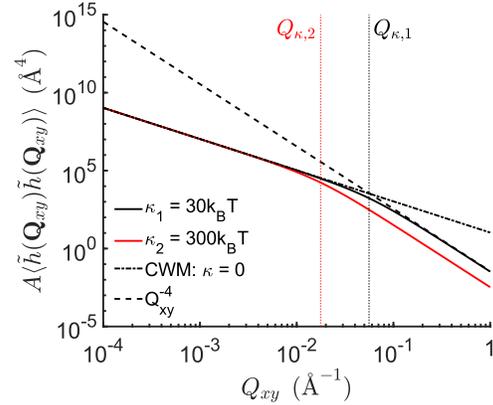

FIG 2. PSDs of the eCWM (area scaled) under two different bending modulus values given in the legend (black and red solid lines), compared to the PSD of the CWM ($\kappa = 0$, dash-dotted lines). They are calculated according to Eq. (1) for $a_m = 5$ Å, $\gamma = 38$ mN/m at 293 K. $Q_{xy}^{-4}$ dependence is entered as a guide-to-the-eye (dashed line) and the two vertical lines depict the wavenumber cut-offs $Q_{\kappa,1} = 1/L_{\kappa,1}$ and $Q_{\kappa,2} = 1/L_{\kappa,2}$ for the two moduli.

*Contact author: chen.shen@desy.de

†Contact author: ocko@bnl.gov

## B. Differential scattering cross section in the extended Capillary Wave Model

The differential scattering cross section $\frac{d\sigma}{d\Omega}$ for the eCWM is computed from the height-height correlation function, $\langle h(\boldsymbol{r}_{xy})h(\boldsymbol{0})\rangle$, which considers the bending modulus (Eq. (2)). For both specular reflectivity and the surrounding diffuse scattering, the differential scattering cross section has the same expression. It is the 3D Fourier transform of $\rho_b(\boldsymbol{r}_{xy}, z) = \rho_{b,0}(z - h(\boldsymbol{r}_{xy}))$. Using the Fourier transform convolution theorem, $\frac{d\sigma}{d\Omega}$ becomes the product of the squared z-direction Fourier transform of the intrinsic SLD profile $\rho_{b,0}(z)$ and the 2D Fourier transform on $h(\boldsymbol{r}_{xy})$ with a phase shift $Q_z[h(\boldsymbol{r}_{xy}) - h(\boldsymbol{r}''_{xy})]$ (derivation shown in APPENDIX B):

$$\frac{d\sigma}{d\Omega} = \frac{\rho_{b,\infty}^2 |t_\alpha|^2 |t_\beta|^2}{Q_z^2} \cdot |\Phi(Q_z)|^2 \cdot \iint \exp\left(iQ_z[h(\boldsymbol{r}_{xy}) - h(\boldsymbol{r}''_{xy})] + i\boldsymbol{Q}_{xy} \cdot (\boldsymbol{r}_{xy} - \boldsymbol{r}''_{xy})\right) d\boldsymbol{r}_{xy} d\boldsymbol{r}''_{xy}$$

$$= A_0 \cdot \left(\frac{Q_c}{2Q_z}\right)^4 |t_\alpha|^2 |t_\beta|^2 \cdot |\Phi(Q_z)|^2 \cdot \psi(Q_{xy}, Q_z)$$

(6)

Here $\boldsymbol{Q}_{xy} = (Q_x, Q_y) = (k_0 \cos\beta \sin 2\theta, k_0(\cos\beta \cos 2\theta - \cos\alpha))$ and its modulus $Q_{xy} = \sqrt{Q_x^2 + Q_y^2}$, and $Q_z = k_0 \cdot (\sin\alpha + \sin\beta)$ (FIG 1). The wavenumber $k_0 = |\boldsymbol{k}_i| = |\boldsymbol{k}_f| = \frac{2\pi}{\lambda}$. $\boldsymbol{r}''_{xy}$ is another position in the surface plane. $A_0$ and $Q_c = \sqrt{16\pi\rho_{b,\infty}}$ are the unit cross section area of the incident beam and the critical angle of the liquid surface in $\boldsymbol{Q}$-space, while $\rho_{b,\infty}$ is the bulk SLD. The aforementioned z-direction Fourier transform of $\rho_{b,0}(z)$ is defined as the intrinsic structure factor $\Phi(Q_z) = \frac{1}{\rho_{b,\infty}} \int_{-\infty}^{\infty} \frac{d\rho_{b,0}(z)}{dz} e^{iQ_z \cdot z} dz$. $|t_\alpha|$ and $|t_\beta|$ are the transmission coefficients for the incident and the scattered waves, respectively. They are close to unity except when, respectively, $\alpha$ or $\beta$ are close to the critical angle [38]. Note that the theory presented in this manuscript applies to both vapor-liquid and liquid-liquid interfaces, and for the latter, $\rho_{b,\infty}$ is replaced by the difference $\Delta\rho_{b,\infty}$ of the SLD between the subphase and the superphase.

We introduce the differential roughness factor:

$$\psi(Q_{xy}, Q_z) = \frac{Q_z^2}{16\pi^2 \sin\alpha} \cdot \exp(-Q_z^2 \langle h^2\rangle) \int \exp(Q_z^2 \langle h(\boldsymbol{r}_{xy}) h(\boldsymbol{0})\rangle) \exp(i\boldsymbol{Q}_{xy} \cdot \boldsymbol{r}_{xy}) d\boldsymbol{r}_{xy}$$

(7)

that fully describes the effect of the surface height profile ($h(\boldsymbol{r}_{xy})$) on the differential scattering cross section (derived in APPENDIX B). In the following, the analytical form of $\psi(Q_{xy}, Q_z)$ is derived for the eCWM. When relating to measured scattered intensities, the regular roughness factor, $\Psi(Q_{xy}, Q_z)$, is used which integrates $\psi(Q_{xy}, Q_z)$ over the solid angle of the detector (introduced in Eq. (16)).

For the eCWM, the height-height correlation in $\psi(Q_{xy}, Q_z)$ is given by Eq. (2). The longest length scale that can be probed in an X-ray measurement is limited by the coherent length $\xi_{coh}$, typically up to some tens of micrometers [39]. Hence the height-height correlation in Eq. (2) can be further simplified by limiting $r'_{xy} \leq \xi_{coh} \ll L_g$ and applying $\lim_{\frac{r'_{xy}}{L_g} \to 0} K_0\left(\frac{r'_{xy}}{L_g}\right) \approx -\ln\left(\frac{0.89 r'_{xy}}{L_g}\right)$. This yields:

$$\langle h(\boldsymbol{r}_{xy})h(\boldsymbol{0})\rangle \approx \frac{k_B T}{2\pi\gamma}\left[-\ln\left(\frac{0.89 r'_{xy}}{L_g}\right) - K_0\left(\frac{r'_{xy}}{L_\kappa}\right)\right]$$

(8)

The $K_0\left(\frac{r'_{xy}}{L_\kappa}\right)$ term cannot be approximated to a logarithmic form since the upper limit of $r'_{xy}$ is $\xi_{coh} > L_\kappa$ and hence $\frac{r'_{xy}}{L_\kappa}$ is not close to 0.

The expression of the differential roughness factor $\psi(Q_{xy}, Q_z)$ for the eCWM is derived by inserting

*Contact author: chen.shen@desy.de

†Contact author: ocko@bnl.gov

$\langle h(r_{xy})h(0)\rangle$ and $\langle h^2\rangle$ from Eq. (8) and Eq. (3) into Eq. (7) [§] (see detailed derivation in APPENDIX C):

$$\psi_{eCWM}(Q_{xy}, Q_z) = \frac{Q_z^4}{16\pi^2 \sin\alpha} \cdot \left(\frac{1}{Q_{max}}\right)^\eta \cdot \exp\left[\eta K_0\left(\frac{1}{L_\kappa Q_{max}}\right)\right] \cdot \left[\frac{\chi(\eta)}{Q_{xy}^{2-\eta}} + \frac{C'(Q_{xy}, \eta, L_\kappa)}{Q_z^2}\right]$$
(9)

The dimensionless parameter $\eta = \frac{k_B T}{2\pi\gamma} Q_z^2$. There are three terms in Eq. (9):

$$\chi(\eta) = \frac{k_B T}{\gamma}\Lambda(\eta)$$

$$\Lambda(\eta) = \frac{1}{\eta} \cdot \frac{2^{1-\eta} \cdot \Gamma\left(1 - \frac{\eta}{2}\right)}{\Gamma\left(\frac{\eta}{2}\right)}$$

$$C'(Q_{xy}, \eta, L_\kappa) = 2\pi \int_0^{8L_\kappa} (r'_{xy})^{1-\eta} \left[\exp\left(-\eta K_0\left(\frac{r'_{xy}}{L_\kappa}\right)\right) - 1\right] J_0(Q_{xy} r'_{xy}) dr_{xy}$$

Note that for $\eta < 1.5$, the first two terms above can be approximated to $\Lambda(\eta) \approx 1$ and $\chi(\eta) \approx \frac{k_B T}{\gamma}$ [40] [**]. $J_0(x)$ is the Bessel function of the 1st kind of the 0th order. The rationalization of the limited integration range of $8L_\kappa$ in the third term is provided in APPENDIX D.

The differential roughness factor $\psi_{eCWM}$ depends on the temperature, tension and bending rigidity through $\eta$ and $L_\kappa$. The systems of interest here are those with a bending rigidity larger than $2k_B T$, and a typical molecular size of 3 ~ 5 Å. For these systems $L_\kappa > 2Q_{max}^{-1}$, and the differential roughness factor can be further simplified to

$$\psi_{eCWM}(Q_{xy}, Q_z) \approx \frac{Q_z^4}{16\pi^2 \sin\alpha} \cdot \left(\frac{1}{0.89 Q_\kappa}\right)^\eta \cdot \left[\frac{\chi(\eta)}{Q_{xy}^{2-\eta}} + \frac{C'(Q_{xy}, \eta, L_\kappa)}{Q_z^2}\right]$$
(10)

by applying $\lim_{\frac{1}{L_\kappa Q_{max}} \to 0} K_0\left(\frac{1}{L_\kappa Q_{max}}\right) \approx -\ln\left(\frac{0.89}{L_\kappa Q_{max}}\right)$ and $Q_\kappa = L_\kappa^{-1}$. Eq. (10) shows that the presence of the rigidity related cut-off $Q_\kappa$ in the eCWM differential roughness factor is analogous to the molecular cut-off $Q_{max}$ in the factor of the CWM [18].

We extend the definition of the differential scattering cross section for the eCWM (Eq. (6)) to explicitly include the temperature, surface tension and bending rigidity as defined by $\psi_{eCWM}$ in Eq. (9):

$$\left(\frac{d\sigma}{d\Omega}\right)_{eCWM} = A_0 \cdot \left(\frac{Q_c}{2Q_z}\right)^4 |t_\alpha|^2 |t_\beta|^2 \cdot |\Phi(Q_z)|^2 \cdot \psi_{eCWM}(Q_{xy}, Q_z)$$

### C. Features of the differential scattering cross section and its approximation forms

We discuss the features of the differential scattering cross section for the eCWM and some useful approximations before proceeding to the specular reflectivity and diffuse scattering calculations that are relevant to experimental results.

The first feature is that in the zero $Q_z$ limit the $Q_{xy}$-dependence of $\frac{d\sigma}{d\Omega}$ must approach that of a PSD (FIG 3(a), see derivation of this property in APPENDIX E). For the eCWM:

$$\lim_{Q_z \to 0} \psi_{eCWM}(Q_{xy}, Q_z) \propto A\langle \tilde{h}(\mathbf{Q}_{xy})\tilde{h}(-\mathbf{Q}_{xy})\rangle$$
$$= \frac{k_B T}{\gamma} \frac{1}{L_g^{-2} + Q_{xy}^2 + L_\kappa^2 Q_{xy}^4}$$
(11)

---

[§] Here the integration up to $|r_{xy}| \to \infty$ implies that the effect of finite $\xi_{coh}$ is not included. This is for simplicity and does not affect the results presented here. The finite $\xi_{coh}$ effect can be introduced by an additional $\exp\left(-\frac{\pi r'^2}{\xi_{coh}}\right)$ term in the integrand: $\frac{d\sigma}{d\Omega} \propto \int_0^\infty (r'_{xy})^{1-\eta} \exp\left(-\eta K_0\left(\frac{r'_{xy}}{L_\kappa}\right)\right) J_0(Q_{xy} r'_{xy}) \exp\left(-\frac{\pi r'^2}{\xi_{coh}}\right) dr_{xy}$ (see Equation 2.33 in Sinha, et. al., Phys Rev B, 1988 and Equation 6.13-14 in Tolan, "X-ray scattering from soft-matter thin films", Springer, 1999).

[**] Tostmann et. al. proposed this equation using the physical meaning of the Q-space integration, despite that it cannot be derived mathematically from the height-height correlation of the CWM.

*Contact author: chen.shen@desy.de

†Contact author: ocko@bnl.gov

At larger $Q_z$, $\psi(Q_{xy}, Q_z)$ deviates from the PSD (FIG 3(b)). Eq. (11) is no longer valid and Eq. (9) must be used. Note that at zero rigidity Eq. (11) reduces to the PSD of the CWM.

The second feature is that $\left(\frac{d\sigma}{d\Omega}\right)_{eCWM}$ converges to the expression for the CWM $\left(\frac{d\sigma}{d\Omega}\right)_{CWM} = \frac{A_0 \rho_{b,\infty}^2 |\Phi(Q_z)|^2}{\sin\alpha} \cdot \left(\frac{1}{Q_{max}}\right)^\eta \cdot \frac{\chi(\eta)}{Q_{xy}^{2-\eta}}$ at the zero rigidity limit. This is shown by calculating the zero-rigidity limit of the eCWM differential roughness factor using the complete form Eq. (9). With zero rigidity, $L_\kappa = 0$ and $Q_\kappa \to \infty$. Since $\lim_{x\to\infty} K_0(x) = 0$, the factor $\exp\left[\eta K_0\left(\frac{Q_\kappa}{Q_{max}}\right)\right] = 1$, and the $C'$ term equals 0. We obtain $\psi_{eCWM}|_{\kappa=0} = \psi_{CWM} = \frac{Q_z^4}{16\pi^2 \sin\alpha} \cdot \left(\frac{1}{Q_{max}}\right)^\eta \cdot \frac{\chi(\eta)}{Q_{xy}^{2-\eta}}$. Note that in the expression for the CWM, $\chi(\eta)$ is often replaced by $\frac{k_B T}{\gamma}$ [18].

The third feature is that $\left(\frac{d\sigma}{d\Omega}\right)_{eCWM}$ can be expressed in terms of a small modification to the CWM differential scattering cross section. This is manifested in separating the differential roughness factor from Eq. (10) into two terms:

$$\psi_{eCWM}(Q_{xy}, Q_z) \approx \underbrace{\frac{Q_z^4}{16\pi^2 \sin\alpha} \cdot \left(\frac{1}{Q_\kappa}\right)^\eta \cdot \frac{\chi(\eta)}{Q_{xy}^{2-\eta}}}_{CWM\ term} + \underbrace{\frac{Q_z^4}{16\pi^2 \sin\alpha} \cdot \left(\frac{1}{Q_\kappa}\right)^\eta \cdot \frac{C'(Q_{xy}, \eta, L_\kappa)}{Q_z^2}}_{C'\ term}$$

(12)

Note that the factor $0.89^{-\eta} \approx 1$ is neglected in the discussion here.

In the following, we show how $\left(\frac{d\sigma}{d\Omega}\right)_{eCWM}$ compares to $\left(\frac{d\sigma}{d\Omega}\right)_{CWM}$ by discussing the $Q_{xy}$ and $Q_z$ dependences of their $\psi$ factors. $\psi_{CWM}$ has a $Q_{xy}^{\eta-2}$ dependence [18]. This dependence implies that $\frac{d\sigma}{d\Omega}$ is singular at the specular reflection position and falls off continuously as $Q_{xy}^{\eta-2}$; the specular reflection and the diffuse scattering around it cannot be separated. Its $\left(\frac{1}{Q_{max}}\right)^\eta$ factor indicates that the shortest wavelength of the thermal capillary waves is on the molecular length scale. $\psi_{eCWM}$ has a similar form as $\psi_{CWM}$, with two modifications (Eq. (12)): (1) a $C'$ term in addition to the $Q_{xy}^{\eta-2}$ dependence; (2) $Q_{max}$ is replaced by $Q_\kappa = \sqrt{\frac{\gamma}{\kappa}}$. The additional term $C'$ is negative (APPENDIX D). Hence it results in a faster fall-off of $\psi_{eCWM}$, and accordingly a faster fall-off in $\frac{d\sigma}{d\Omega}$ compared to the $Q_{xy}^{\eta-2}$ fall-off of the CWM (FIG 3). The negative deviation of $\psi_{eCWM}$ from the $Q_{xy}^{\eta-2}$ fall-off of the CWM is directly attributed to the stiffness $\kappa$ of the film that affects the value of $C'$: the diffuse scattering at large $Q_{xy}$ (short in-plane length scale) is reduced since surface thermal roughness is suppressed by the stiffness. FIG 3 shows that the $Q_{xy}$ dependence of the negative $C'$ term is much flatter compared to the $Q_{xy}^{\eta-2}$ fall-off of the CWM term. As the consequence of the flat $Q_{xy}$ dependence of the $C'$ term, at small $Q_{xy} < \frac{Q_\kappa}{\pi}$, the $C'$ term is relatively small compared to the CWM term, such that $\frac{d\sigma}{d\Omega}$ behaves similarly as with the CWM. This criterion is proposed in our previous publication [31]. At larger $Q_{xy}$, the $C'$ term becomes comparable to the CWM term, and therefore significantly suppresses the amplitude of $\left(\frac{d\sigma}{d\Omega}\right)_{eCWM}$ compared to that of the CWM. Not only is this faster fall-off beyond $Q_{xy}^{\eta-2}$ consistent with the PSD, it was previously observed and reported as a $Q_{xy}^{-3.3}$ dependence in some early diffuse scattering studies [28]. The cross-over to the $Q_{xy}^{\eta-2}$ dependence is found at about $Q_{xy} \approx Q_\kappa$ (FIG 3). As an example, the cross-over occurs at $Q_{xy} = 0.05$ Å$^{-1}$ for a film with $\kappa = 30 k_B T$ at a tension of 38 mN/m. This cross-over point is consistent with the PSD (FIG 2). The second modification is the replacement of $Q_{max} = \frac{\pi}{a_m}$ by $Q_\kappa$. Consequently, $\left(\frac{d\sigma}{d\Omega}\right)_{eCWM}$ falls off more slowly with $Q_z$ compared to $\left(\frac{d\sigma}{d\Omega}\right)_{CWM}$ since $\left(\frac{1}{Q_\kappa}\right)^\eta > \left(\frac{1}{Q_{max}}\right)^\eta$. Moreover, this replacement is equivalent to assuming the shortest wavelength of the excitable capillary wave to be $L_\kappa$, i.e. the largest excitable wavenumber being $Q_\kappa$, and this is consistent with the effect of rigidity on the PSD (FIG 2).

*Contact author: chen.shen@desy.de

†Contact author: ocko@bnl.gov

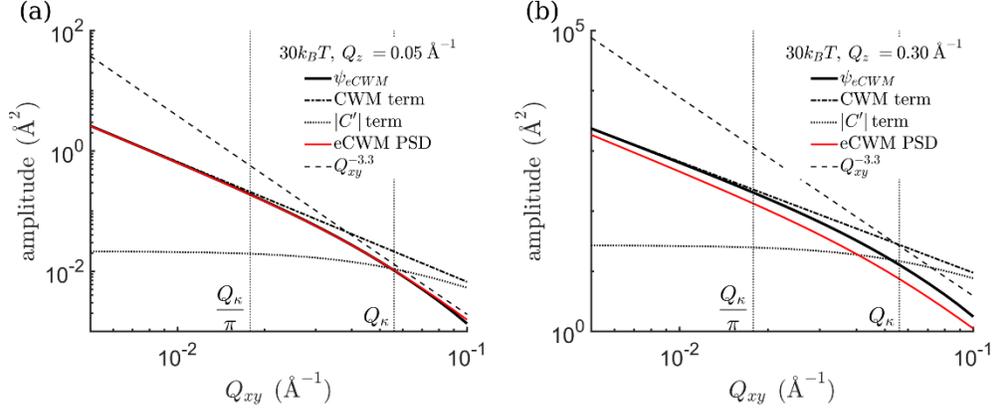

FIG 3. Log-log plots of the $Q_{xy}$-dependence of the differential roughness factor for the eCWM (Eq. (9)), its CWM term and the absolute value of $C'$ term (Eq. (12)) at $Q_z = 0.05$ Å$^{-1}$ (a) and 0.3 Å$^{-1}$ (b). They are compared to the PSD of the eCWM (Eq. (1)) and a $Q_{xy}^{-3.3}$ dependence [28]. Curves are calculated for $\gamma$ = 38 mN/m, $\kappa$ = 30k$_B$T at 293 K, and are normalized by the prefactor $\frac{1}{16\pi^2 \sin\alpha}$. Vertical lines mark the cut-off positions at $\frac{Q_\kappa}{\pi}$ and $Q_\kappa$, as discussed in the text. Note that $\lim_{Q_z \to 0} \psi$ coincides with the PSD.

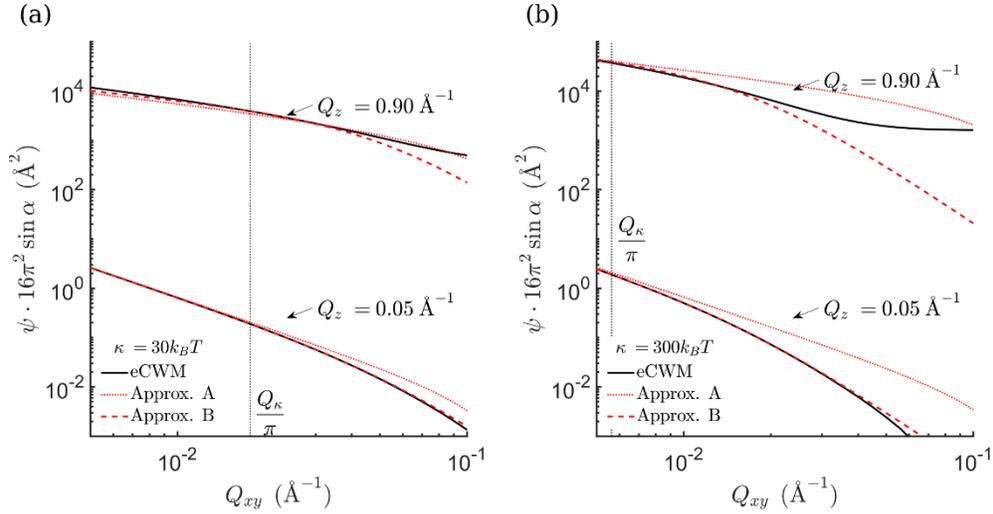

FIG 4. Log-log plots of the $Q_{xy}$-dependence of the eCWM differential roughness factor $\psi_{eCWM}$ compared to the Approx. A (Eq. (13), proposed in ref. [13]) and Approx. B (Eq. (15), proposed in ref. [26]) at $Q_z = 0.05$ and 0.9 Å$^{-1}$. Curves are calculated for two bending moduli: 30k$_B$T (a) and 300k$_B$T (b), at $\gamma$ = 38 mN/m and 293 K. The curves are normalized by the prefactor $\frac{1}{16\pi^2 \sin\alpha}$. As discussed in the section II-C, Approx. A deviates positively from $\psi$, typically when $Q_{xy} > \frac{Q_\kappa}{\pi}$ (vertical dotted line), while Approx. B agrees with $\psi$ at small $Q_z$ and disagree at large $Q_z$.

Two approximations have been introduced in the previous studies to simplify the calculation of the differential scattering cross section from the PSD of the eCWM [13,26]. One commonly used form, denoted here as Approx. A, approximates the height-height correlation (Eq. (2)) as [13,25]:

$$\frac{\langle h(\mathbf{r}_{xy})h(\mathbf{0})\rangle}{\langle h^2\rangle} = \begin{cases} 1, & r_{xy} \leq L_\kappa \\ 1 - \left[\frac{\ln\left(\frac{r_{xy}}{L_\kappa}\right)}{\ln\left(\frac{L_g}{L_\kappa}\right)}\right], & L_\kappa < r_{xy} < L_g \\ 0, & r_{xy} \geq L_g \end{cases}$$

(13)

This approximation (see ref. [13]) implies an assumption that $r_{xy} \gg L_\kappa$, i.e. $\frac{r'_{xy}}{L_\kappa} \to \infty$, the property $\lim_{x \to \infty} K_0(x) = 0$ of the modified Bessel function can be applied such that $\langle h(\mathbf{r}_{xy})h(\mathbf{0})\rangle \approx \frac{k_B T}{2\pi\gamma} K_0\left(\frac{r'_{xy}}{L_g}\right)$. Then the approximation $\lim_{x \to 0} K_0(x) \approx -\ln x$ is utilized for $K_0\left(\frac{r'_{xy}}{L_g}\right)$ since $r_{xy} \ll L_g$, i.e. $\frac{r'_{xy}}{L_g} \approx \frac{r_{xy}}{L_g} \to 0$, such that $\langle h(\mathbf{r}_{xy})h(\mathbf{0})\rangle \approx \frac{k_B T}{2\pi\gamma}\ln\left(\frac{L_g}{r_{xy}}\right)$. Compared to the original form, Approx. A differs in the small length region near $L_\kappa$ where this approximation of $K_0\left(\frac{r'_{xy}}{L_\kappa}\right) \approx 0$ is not valid. Consequently, $\frac{d\sigma}{d\Omega}$ calculated from Approx. A deviates positively from the true $\frac{d\sigma}{d\Omega}$ at large $Q_{xy}$, e.g. when $Q_{xy} > \frac{Q_\kappa}{\pi}$ (FIG 4, dotted red lines). Using this form of $\frac{d\sigma}{d\Omega}$ to analyze the diffuse scattering results in an overestimation of the bending modulus. It is worth noting that this approximation predicts the correct roughness factor for the CWM regime where $Q_{xy}$ is small ($Q_{xy} \ll \frac{Q_\kappa}{\pi}$), hence is applicable for reflectivity (Equation A18-19 of ref. [13]).

The other approximation, denoted here as Approx. B, was proposed in ref. [26], based on a modified Taylor expansion (see derivation in APPENDIX E) [††]:

$$\psi(Q_{xy}, Q_z) \approx \frac{Q_z^4}{16\pi^2 \sin\alpha} \cdot \exp\left[-Q_z^2 \int_{Q_{xy}}^{Q_{max}} \frac{A\langle \tilde{h}(\mathbf{Q}'_{xy})\tilde{h}(-\mathbf{Q}'_{xy})\rangle}{2\pi} Q'_{xy} dQ'_{xy}\right] \cdot \left[A\langle \tilde{h}(\mathbf{Q}_{xy})\tilde{h}(-\mathbf{Q}_{xy})\rangle\right]$$

(14)

and the differential roughness factor is approximated as:

$$\psi_{eCWM}(Q_{xy}, Q_z) \approx \frac{Q_z^4}{16\pi^2 \sin\alpha} \cdot \left(\frac{1}{Q_\kappa}\right)^\eta \cdot \frac{k_B T}{\gamma} \frac{Q_{xy}^\eta}{Q_{xy}^2 + L_\kappa^2 Q_{xy}^4}$$

(15)

This expression is still only applicable to a limited $Q_z$ range (~ up to 0.6 Å$^{-1}$) due to the nature of the Taylor expansion approximation, despite that the modification (see APPENDIX E) enlarges its valid range beyond that of a standard Taylor expansion. This is shown in FIG 4 from the comparison of this Approx. B (Eq. (15)) with the accurate form (Eq. (9)) at small and at large $Q_z$. Hence Approx. B given by Eq. (15) can be conveniently used to determine the bending rigidity from the $Q_{xy}$ dependence of the diffuse scattering in the low $Q_z$ range, leveraging its computation simplicity, while Eq. (9) must be used to analyze the diffuse scattering data for the whole $Q_z$ range. Note that for completeness, we use Eq. (9) in later sections unless noted otherwise.

We also note that Eq. (15) does not allow a smooth transition to the CWM expression since $\left(\frac{1}{Q_\kappa}\right)$ goes to zero in the zero-rigidity limit. The smooth transition can be achieved by replacing $\left(\frac{1}{Q_\kappa}\right)^\eta$ by $\left(\frac{1}{Q_{max}}\right)^\eta e^{\eta K_0\left(\frac{1}{L_\kappa Q_{max}}\right)}$:

$$\psi_{eCWM}(Q_{xy}, Q_z) \approx \frac{Q_z^4}{16\pi^2 \sin\alpha} \cdot \left(\frac{1}{Q_{max}}\right)^\eta \cdot \exp\left[\eta K_0\left(\frac{1}{L_\kappa Q_{max}}\right)\right] \cdot \frac{k_B T}{\gamma} \frac{Q_{xy}^\eta}{Q_{xy}^2 + L_\kappa^2 Q_{xy}^4}$$

This follows a similar treatment provided in Eq. (10) from the accurate expression given in Eq. (9).

### D. Specular reflectivity and diffuse scattering under the extended Capillary Wave Model

For the specular reflection and its diffuse scattering, their intensity $I$ normalized by the incident intensity $I_0$ are both an integral of the differential cross section around their corresponding scattering angle ($\beta, 2\theta$) over angular openings $\Delta\beta$ and $\Delta 2\theta$ [31,41]:

---

[††] Note that the Equation 3 of Ref. 27, misses the factor $Q_z^2$, and it uses a different PSD expression that does not contain the film area A.

*Contact author: chen.shen@desy.de

†Contact author: ocko@bnl.gov

$$R, R^* = \frac{I}{I_0} = \frac{1}{A_0} \int_{\Delta\beta, \Delta 2\theta} \frac{d\sigma}{d\Omega} d\Omega$$

$$= \left(\frac{Q_c}{2Q_z}\right)^4 |t_\alpha|^2 |t_\beta|^2 \cdot |\Phi(Q_z)|^2 \cdot \Psi(Q_{xy}, Q_z)$$

$$\Psi(Q_{xy}, Q_z) = \iint_{\Delta\beta, \Delta 2\theta} \psi(Q_{xy}, Q_z) d(\sin\beta) d2\theta$$

(16)

where $d\Omega = d(\sin\beta) d2\theta$. The roughness factor $\Psi(Q_{xy}, Q_z)$ is defined for the integral of the differential roughness factor $\psi(Q_{xy}, Q_z)$. This expression fully describes the $Q_{xy}$ and $Q_z$ dependences of the specular reflectivity and the diffuse scattering intensity that originates from the thermal roughness. Note that the operation of moving $Q_z^{-4}$, $|t_\beta|^2$ and $|\Phi(Q_z)|^2$ out of the integral is valid for small $\Delta\beta$ such that these three factors can be considered invariant within $\Delta\beta$.

Specular reflectivity $R$ corresponds to the case when the integral in the roughness factor is centered around the specular condition ($\alpha = \beta$, $2\theta = 0°$) with full-width-at-half-maximum (FWHM) angular openings $\Delta\beta$ and $\Delta 2\theta$. This corresponds to $2\theta$ ranging from $-\Delta 2\theta/2$ to $+\Delta 2\theta/2$, and to $\beta$ ranging from $\alpha - \Delta\beta/2$ to $\alpha + \Delta\beta/2$ (FIG 1) [18]. The angular openings give an in-plane $Q_{xy}$-resolution $\delta Q_{xy,R}$ (half-width-at-half-maximum HWHM) [‡‡] [31] and the reflectivity under this resolution is:

$$R = \left(\frac{Q_c}{2Q_z}\right)^4 |t_\alpha|^4 \cdot |\Phi(Q_z)|^2 \cdot \Psi_R(Q_z)$$
$$\approx R_F |\Phi(Q_z)|^2 \cdot \Psi_R(Q_z)$$

$$\Psi_R(Q_z) = \int\!\!\!\int_{\alpha-\frac{\Delta\beta}{2}, -\frac{\Delta 2\theta}{2}}^{\alpha+\frac{\Delta\beta}{2}, +\frac{\Delta 2\theta}{2}} \psi(Q_{xy}, Q_z) d(\sin\beta) d2\theta$$

(17)

$R_F \equiv \left|\frac{Q_z - \sqrt{Q_z^2 - Q_c^2}}{Q_z + \sqrt{Q_z^2 - Q_c^2}}\right|^2 \approx \left(\frac{Q_c}{2Q_z}\right)^4 |t_\alpha|^4$ is the Fresnel reflectivity of an ideally flat and sharp interface. For $\alpha$ larger than twice the critical angle $\alpha_c = \operatorname{asin}\frac{Q_c \lambda}{4\pi}$, $R_F \approx \left(\frac{Q_c}{2Q_z}\right)^4$ since $|t_\alpha|$ is close to unity.

We will first evaluate the contributions to the eCWM roughness factor, for all scattering conditions, i.e. both the specular reflection $R$ and the surrounding diffuse scattering $R^*$. This roughness factor contains all the effects of its thermal roughness on the measured scattering signal. With $\psi_{eCWM}$ given by Eq. (9), the eCWM roughness factor is:

$$\Psi_{eCWM}(Q_{xy}, Q_z) = \iint_{\Delta\beta, \Delta 2\theta} \psi_{eCWM}(Q_{xy}, Q_z) d(\sin\beta) d2\theta$$
$$= \frac{Q_z^4}{16\pi^2 \sin\alpha} \cdot \left(\frac{1}{Q_{max}}\right)^\eta \cdot \exp\left[\eta K_0 \left(\frac{1}{L_\kappa Q_{max}}\right)\right]$$
$$\cdot \left[\chi(\eta) \iint_{\Delta\beta, \Delta 2\theta} \frac{1}{Q_{xy}^{2-\eta}} d(\sin\beta) d2\theta + \frac{1}{Q_z^2} \iint_{\Delta\beta, \Delta 2\theta} C'(Q_{xy}, \eta, L_\kappa) d(\sin\beta) d2\theta\right]$$

(18)

Note that this roughness factor depends on the temperature, tension and the bending rigidity through $\eta$ and $L_\kappa$, like the differential factor $\psi_{eCWM}$. Moreover, $\Delta\beta$ in practice is usually small enough such that $Q_z$ can be considered invariant within $\Delta\beta$.

An important feature that emerges from the roughness factor expression is that the surface thermal roughness induces diffuse scattering (around the specular reflection). Similar to the CWM, the eCWM diffuse scattering cannot be distinguished from the specular reflection [18]. Approaching the specular position ($Q_{xy} \to 0$), line shape of the scattering signal continues as $\frac{1}{Q_{xy}^{2-\eta}}$ towards a singularity ($C'$ only contributes to large $Q_{xy}$; see above). Consequently, angular opening ($\Delta\beta$, $\Delta 2\theta$) around the specular position always includes remnants of the diffuse scattering, and hence the reflectivity fall-off depends on the in-plane resolution $\delta Q_{xy,R}$ defined by the angular opening.

---

[‡‡] $(\delta Q_{x,R}, \delta Q_{y,R}) = k_0 \cdot \left(\frac{\Delta 2\theta}{2}, \sin\beta\left(\frac{\Delta\beta}{2}\right)\right)$. Mind that $\delta Q_{xy,R}$ is a HWHM while the angular openings $\Delta 2\theta$ and $\Delta\beta$ are FWHM.

*Contact author: chen.shen@desy.de

†Contact author: ocko@bnl.gov

In order to analytically evaluate the effect of the eCWM thermal roughness on the reflectivity it is convenient to assume a circular $Q_{xy}$-resolution, i.e. with a single $\delta Q_{xy,R}$-value which is a radius in $Q$-space. This corresponds to an elliptical angular opening as $\Delta 2\theta \ll \Delta\beta$ ‡‡. Under such a resolution $\iint(...)dQ_x dQ_y = 2\pi \int(...)Q_{xy}dQ_{xy}$ and $d(\sin\beta)d2\theta \approx \frac{dQ_x dQ_y}{k_0^2 \sin\alpha}$ are applied to both integral terms in the roughness factor (Eq. (18)) [§§], and applying $\lim_{Q_{xy}\to 0} J_0(Q_{xy}r_{xy}') = 1$ to the 2nd term yields the roughness factor expression under the circular resolution:

$$\Psi_{R,eCWM}(Q_z)\big|_{\delta Q_{xy,R}} = \int_0^{\delta Q_{xy,R}} \psi_{eCWM}(Q_{xy},Q_z)d\Omega$$
$$= \left(\frac{\delta Q_{xy,R}}{Q_{max}}\right)^\eta \cdot \exp\left[\eta K_0\left(\frac{1}{L_\kappa Q_{max}}\right)\right] \cdot \left[\Lambda(\eta) + \frac{\delta Q_{xy,R}^{2-\eta}}{4\pi} \cdot C(\eta,L_\kappa)\right] \quad (19)$$

where $C(\eta,L_\kappa) = 2\pi \int_0^{8L_\kappa}(r_{xy}')^{1-\eta}\left[\exp\left(-\eta K_0\left(\frac{r_{xy}'}{L_\kappa}\right)\right) - 1\right]dr_{xy}$

And the specular reflectivity under the circular resolution is:

$$R(Q_z)\big|_{\delta Q_{xy,R}} = R_F(Q_z) \cdot |\Phi(Q_z)|^2 \cdot \Psi_{R,eCWM}(Q_z)\big|_{\delta Q_{xy,R}} \quad (20)$$

Similar to the CWM case [18], the eCWM roughness factor results in a $\delta Q_{xy,R}$ dependent fall-off of the reflectivity with $Q_z$, in addition to the fall-off of $R_F(Q_z)|\Phi(Q_z)|^2$ with $Q_z$ (FIG 5(a), black solid line vs unity). This fall-off is slower compared to the CWM ($\kappa = 0$).

Similar to the treatment used in the CWM [18], the eCWM roughness factor can be parameterized using a Gaussian factor with an eCWM thermal roughness $\sigma_{R,eCWM}$:

$$\Psi_{R,eCWM}(Q_z) = \frac{R(Q_z)}{R_F(Q_z)|\Phi(Q_z)|^2} \equiv \exp(-Q_z^2 \sigma_{R,eCWM}^2)$$
$$\sigma_{R,eCWM}^2 = -\frac{1}{Q_z^2}\ln\left(\Psi_{R,eCWM}(Q_z)\right) \quad (21)$$

Note that this expression for $\sigma_{R,eCWM}^2$ applies not only to a circular resolution, but also to any resolution function $(\Delta\beta,\Delta 2\theta)$ through the angular integration defined in Eq. (17). Hence the circular resolution subscript is dropped.

With this parameterization, the statistically and time averaged SLD profile $\rho_{b,eCWM}(z)$ over an in-plane length scale that corresponds to the in-plane resolution of the reflectivity is modelled by an error function (erf($z$)) with the rms width $\sigma_{R,eCWM}$:

$$\rho_{b,eCWM}(z) = \frac{\rho_{b,\infty}}{2}\left(1 + \text{erf}\left(\frac{z}{\sqrt{2}\sigma_{R,eCWM}}\right)\right) \quad (22)$$

For a circular resolution function, the roughness $\sigma_{R,eCWM}$ is obtained by inserting $\Psi_{R,eCWM}(Q_z)\big|_{\delta Q_{xy,R}}$ from Eq. (19) into Eq. (21):

$$\sigma_{R,eCWM}^2 = \frac{k_B T}{2\pi\gamma}\ln\frac{Q_{max}}{\delta Q_{xy,R}} - \frac{k_B T}{2\pi\gamma}K_0\left(\frac{Q_\kappa}{Q_{max}}\right)$$
$$-\frac{1}{Q_z^2}\ln\left(\Lambda(\eta) + \frac{\delta Q_{xy,R}^{2-\eta}}{4\pi}C(\eta,L_\kappa)\right) \quad (23)$$

For clarity, we use $Q_\kappa$ rather than $L_\kappa$ in the $K_0$ function. The first term is the full expression for the CWM roughness ($\kappa = 0$) and the last two terms depend on the bending modulus $\kappa$. To evaluate the effect of $\kappa$, we consider values of $\kappa$ larger than $2k_B T$, and a typical molecular size of 3 ~ 5 Å, such that $Q_{max} > 2Q_\kappa$. This allows one to apply $\lim_{x\to 0} K_0(x) \approx -\ln(0.89x)$ and yields the expression for $\sigma_{R,eCWM}^2$ as the sum of two terms:

---

[§§] Here two equalities are applied: (1) $dQ_x dQ_y = Q_{xy} dQ_{xy} d2\theta'$; (2) $dQ_x dQ_y = (k_0 \cos\beta \cos 2\theta\, d2\theta) \cdot (k_0 \sin\beta \cos 2\theta\, d\beta) \approx k_0^2 \sin\alpha\, d(\sin\beta)d2\theta$ under the conditions $\cos 2\theta \approx 1$ for small $2\theta$, and $\alpha = \beta$.

*Contact author: chen.shen@desy.de

†Contact author: ocko@bnl.gov

$$\sigma_{R,eCWM}^2 = \underbrace{\frac{k_BT}{2\pi\gamma}\ln\frac{0.89Q_\kappa}{\delta Q_{xy,R}}}_{\text{term 1}} - \underbrace{\frac{1}{Q_z^2}\ln\left(\Lambda(\eta) + \frac{\delta Q_{xy,R}^{2-\eta}}{4\pi}C(\eta,L_\kappa)\right)}_{\text{term 2}} \quad (24)$$

Term (1) is analogous to the CWM expression: $Q_{max}$ in the CWM is replaced by the stiffness dependent cut-off $0.89Q_\kappa$. This is the dominant term except at very large $Q_z$ where term (2) is relevant, as illustrated by the deviation of the blue and black lines in FIG 5(a). This is the case since $\Lambda(\eta)$ is close to unity except when $\eta$ approaches 2 (in the scattering theory for liquid surfaces $\eta < 2$ ). Its second component, $\frac{\delta Q_{xy,R}^{2-\eta}}{4\pi}C(\eta,L_\kappa)$, is negligible compared to $\Lambda(\eta)$ for a bending rigidity $\kappa < 100k_BT$ (APPENDIX D). In the range of $2k_BT < \kappa < 100k_BT$, that is common for soft matter thin films, the expressions for the specular reflectivity and the thermal roughness are well approximated by:

$$R(Q_z) \approx R_F(Q_z) \cdot |\Phi(Q_z)|^2 \cdot \left(\frac{\delta Q_{xy,R}}{0.89Q_\kappa}\right)^\eta$$
$$\sigma_{R,eCWM}^2 \approx \frac{k_BT}{2\pi\gamma}\ln\frac{0.89Q_\kappa}{\delta Q_{xy,R}} \quad (25)$$

This expression of the thermal roughness for the eCWM is consistent with the Eq. A19 given in ref. [13] that is derived from Approx. A of the correlation function (Eq. (13)), except for the effect of the resolution, discussed in the following paragraphs.

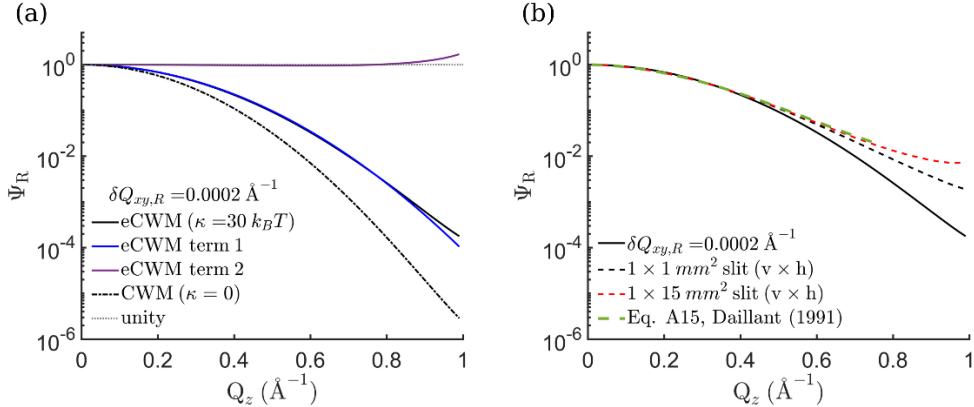

FIG 5. $Q_z$-dependence of the eCWM roughness factor $\Psi_R$ for $\gamma = 38$ mN/m, $T = 293$ K, $\kappa = 30k_BT$ and $a_m = 5$ Å. (a) $\Psi_R$ calculated using a circular resolution function ($\delta Q_{xy,R} = 2\times 10^{-4}$ Å$^{-1}$) is shown with its two contributing terms to $\Psi_R$ (defined in Eq. (24)), and is compared to the $\Psi_R$ of the CWM ($\kappa = 0$) and a value of unity. (b) $\Psi_R$ calculated using the circular resolution is compared to its values using two different rectangular slit resolutions (with a 1 m sample to slit distance at 15 keV), and to the expression defined in Equation A15 in ref. [13] for 1 mm vertical slit [***]. Note that the 1×1 mm$^2$ slit corresponds to $\delta Q_{y,R} = 2\times 10^{-4}$ Å$^{-1}$ at $Q_z = 0.8$ Å$^{-1}$ and $\delta Q_{x,R} = 3.8\times 10^{-3}$ Å$^{-1}$ at all $Q_z$. As discussed in the text, the Eq. A15 from ref. [13] corresponds to an infinitely wide horizontal slit.

As it is discussed in our earlier publication [31], a circular resolution function does not accurately represent the resolution function of a standard reflectometer. An accurate representation is a rectangular resolution function (FIG 1) as it relates to the detector slit geometry or the rectangular region of interest from a 2D detector. Accordingly, $\delta Q_{x,R} \gg \delta Q_{y,R}$, and $\delta Q_{y,R}$ increases with $\beta$ [‡‡]. In this case there is no analytical solution like the circular resolution case, and the reflectivity must be calculated by a numerical integral over its rectangular resolution using Eq. (18). The thermal roughness induced fall-off can be evaluated by inserting the result from Eq. (18) into Eq. (21). FIG 5(b) shows the reflectivity fall-off under two different rectangular resolution conditions and is compared to that using a circular resolution $\delta Q_{xy,R} = 2\times 10^{-4}$ Å$^{-1}$. These two slit resolutions have the same $\Delta\beta$ but different $\Delta 2\theta$ opening. In the following, we use a $\Delta\beta$ corresponding to $\delta Q_{y,R} = 2\times10^{-4}$ Å$^{-1}$ at $Q_z = 0.8$ Å$^{-1}$, being the same resolution value as the circular resolution. The maximal position of $Q_z$, typically close to 0.8 Å$^{-1}$, is where the $\delta Q_{y,R}$ resolution is the widest and that is why we picked this value. The $\delta Q_{x,R}$ is always much wider than this maximal $\delta Q_{y,R}$. The reflectivity fall-off is less at high $Q_z$ compared to that of the circular resolution and this is due to the wider $\delta Q_{x,R}$.

---

[***] The x-direction in Daillant 1991 is the longitudinal direction, i.e. y-direction in this paper.

*Contact author: chen.shen@desy.de

†Contact author: ocko@bnl.gov

Our reflectivity expression using the roughness factor of the eCWM from Eq. (18) is consistent with the expression proposed by the pioneering studies for rigid films on liquid surfaces (Eq. A15 and A19 in ref. [13]). Their expression corresponds to a fully open slit in the x-direction, i.e. $\delta Q_{x,R} \to \infty$. Note that this full integration along the x direction results in a mathematical limit of $\eta \leq 1$ which limits the applicable $Q_z$ range of their expression. The resulting curve is nearly the same as that derived from Eq. (18) using a very large horizontal slit, e.g. 15 mm wide (FIG 5(b), green dashed line vs red dashed line), except that Eq. (18) allows for the full $Q_z$ range as long as $\eta \leq 2$.

The expression of the diffuse scattering $R^*$ appears nearly the same as for the specular reflection expression except for two differences: (1) the integration in the roughness factor $\Psi_{\text{eCWM}}$ (Eq. (18)) is around an off-specular angle, i.e. $2\theta \neq 0$ and $\beta \neq \alpha$, and (2) a difference in the values of $|t_\alpha|$ and $|t_\beta|$ when $\alpha \neq \beta$:

$$R^*(Q_z, Q_{xy}) = \left(\frac{Q_c}{2Q_z}\right)^4 |t_\alpha|^2 |t_\beta|^2 \cdot |\Phi(Q_z)|^2 \cdot \Psi_{\text{eCWM}}(Q_{xy}, Q_z) \quad (26)$$

This equation allows one to extract the bending modulus directly by the $Q_{xy}$-dependence of the diffuse scattering when the temperature and the surface tension are both known. Thereafter, the intrinsic structure factor $|\Phi(Q_z)|^2$ can be computed as:

$$|\Phi(Q_z)|^2 = \frac{R^*(Q_z, Q_{xy})}{\Psi_{\text{eCWM}}(Q_{xy}, Q_z)} \cdot \frac{(2Q_z)^4}{Q_c^4 |t_\alpha|^2 |t_\beta|^2} \quad (27)$$

### E. Pseudo reflectivity method

Similar to the CWM [31], the combination of Eq. (26) with Eq. (20) provides a quantitative expression of the ratio $r$ for the eCWM between the diffuse scattering $R^*$ and its reflectivity $R|_{\delta Q_{xy,R}}$ under a given in-plane resolution $\delta Q_{xy,R}$:

$$r(Q_z, Q_{xy})|_{\delta Q_{xy,R}} = \frac{R^*(Q_{xy}, Q_z)}{R(Q_z)|_{\delta Q_{xy,R}}} = \frac{\left(\frac{Q_c}{2Q_z}\right)^4 |t_\alpha|^2 |t_\beta|^2}{R_F(Q_z)} \cdot \frac{\Psi_{\text{eCWM}}(Q_{xy}, Q_z)}{\Psi_{R,\text{eCWM}}(Q_z)}$$

$$\approx \frac{Q_c^4 |t_\alpha|^2 |t_\beta|^2}{R_F(Q_z) \cdot (16\pi)^2 \sin\alpha} \cdot \frac{1}{\Lambda(\eta)\delta Q_{xy,R}^\eta} \cdot \left[\chi(\eta) \iint\limits_{\Delta\beta, \Delta 2\theta} \frac{1}{Q_{xy}^{2-\eta}} d(\sin\beta)d2\theta + \frac{1}{Q_z^2} \iint\limits_{\Delta\beta, \Delta 2\theta} C'(Q_{xy}, \eta, L_\kappa) d(\sin\beta)d2\theta\right] \quad (28)$$

Here $\frac{\delta Q_{xy,R}^{2-\eta}}{4\pi} C(\eta, L_\kappa) \ll \Lambda(\eta)$ for specular reflectivity and is ignored (APPENDIX D). For a measurement at sufficiently large $2\theta$ compared to the width $\Delta 2\theta$, for example $2\theta > 1.5 \times \Delta 2\theta$, the value of $Q_{xy}$ can be considered invariant within $2\theta \pm \frac{\Delta 2\theta}{2}$. Moreover, for $Q_z > 3Q_c$, $R_F \approx \left(\frac{Q_c}{2Q_z}\right)^4$ and $|t_\beta|^2 \approx 1$. The integration can be simplified to:

$$r(Q_z, Q_{xy})|_{\delta Q_{xy,R}} \approx \frac{|t_\alpha|^2 Q_z^4}{16\pi^2 \sin\alpha} \cdot \frac{\frac{\chi(\eta)}{Q_{xy}^{2-\eta}} + \frac{C'(Q_{xy}, \eta, L_\kappa)}{Q_z^2}}{\Lambda(\eta)\delta Q_{xy,R}^\eta} \cdot \Delta 2\theta \Delta\beta \quad (29)$$

Here $r(Q_z, Q_{xy})|_{\delta Q_{xy,R}}$ allows one to calculate the reflectivity curve for a given $\delta Q_{xy,R}$ from the diffuse scattering data, e.g. measured by GIXOS [31,42]:

*Contact author: chen.shen@desy.de

†Contact author: ocko@bnl.gov

$$R_{pseudo}(Q_z)\big|_{\delta Q_{xy,R}} = \frac{R^*(Q_z, Q_{xy})}{r(Q_z, Q_{xy})\big|_{\delta Q_{xy,R}}} \quad (30)$$

The reflectivity obtained in this way from the diffuse scattering is referred to as pseudo reflectivity, $R_{pseudo}$, in order to distinguish it from the reflectivity curve measured from conventional reflectometry along the specular axis [31].

The $\frac{c'(Q_{xy},\eta,L_\kappa)}{Q_z^2}$ term is negligible for $Q_{xy} < \frac{Q_\kappa}{\pi}$. In this case, Eq. (29) is simplified to the CWM expression [31]:

$$r\big|_{Q_{xy}<\frac{Q_\kappa}{\pi}} \approx \frac{|t_\alpha|^2}{16\pi^2 \sin\alpha} \cdot \frac{k_B T}{\gamma} \cdot \frac{Q_z^4}{Q_{xy}^{2-\eta} \delta Q_{xy,R}^\eta} \cdot \Delta 2\theta \Delta\beta \quad (31)$$

In summary, Eq. (20) and Eq. (26) describe the specular reflectivity and diffuse scattering of the eCWM for a liquid interface with a thin film at the interface, while the Eq. (28) and Eq. (29) provide their ratio. At $\kappa > 2k_B T$, the specular reflectivity expression is simplified to Eq. (25). Eq. (30) makes it possible to calculate the pseudo specular reflectivity from the diffuse scattering, as is also the case for the CWM [31]. At small $Q_{xy}$, the ratio (Eq. (29)) used in the Eq. (30) simplifies to the CWM form (Eq. (31)).

### III. Experimental and instrumental details

Experiments were conducted on five phospholipid monolayers deposited on the surface of water or aqueous buffer solutions. These are 1-palmitoyl-2-oleoyl-*sn*-glycero-3-phosphatidylcholine (POPC), 1,2-distearoyl-*sn*-glycero-3-PC (DSPC), 1,2-dipalmitoyl-*sn*-glycero-3-PC (DPPC), 1,2-dipalmitoyl-*sn*-glycero-3-phophatidylglycerol (DPPG), and 1,2-dipalmitoyl-*sn*-glycero-3-phosphoethanolamine (DPPE), all used as purchased from Avanti Polar Lipid. Conditions are given in TABLE III. In addition, scattering from the pure water surface was measured as a reference. The PCs were first dissolved in chloroform (HPLC Plus, $\geq$ 99.9%, amylenes as stabilizer, Sigma-Aldrich Chemie GmbH, Germany) while the other lipids were dissolved in 1:9 (vol/vol) mixture of methanol (GC, 99.9%, Sigma-Aldrich) and chloroform. Pure water (resistivity > 18.2 M$\Omega\cdot$cm at 25°C, total organic carbon < 2 ppb) was obtained from Purelab Ultra system (ELGA LabWater).

TABLE III. Conditions of the lipid monolayers for the X-ray experiments. The surface pressure $\Pi = \gamma_0 - \gamma$ where the surface tension of water $\gamma_0$ = 73 mN/m (20°C).

| film | subphase | T [°C] | $\gamma$ [mN/m] | $\Pi$ [mN/m] |
|---|---|---|---|---|
| POPC | water | 20 | 38 | 35 |
| DPPC | water | 22 | 38 | 35 |
| DPPG | 10 mM tris•HCl, 0.05mM Na$_2$EDTA, pH 7.3 | 22 | 38, 28 | 35, 45 |
| DPPE | 10 mM tris•HCl, 0.05mM Na$_2$EDTA, pH 7.3 | 22 | 38, 28 | 35, 45 |
| DSPC | water | 20 | 28 | 45 |

The X-ray measurements were performed at the beamline P08 of PETRA III (DESY, Hamburg, Germany) [43] and at the Open Platform Liquids Scattering (OPLS) endstation at beamline 12ID-SMI at the National Synchrotron Light Source II (NSLS-II, Brookhaven National Laboratory, USA). The instrumental settings employed are mostly identical to an earlier publication [31] where only the key parameters and those settings which differ from ref. [31] are provided here. GIXOS measurements were conducted using the Langmuir trough grazing incidence diffraction setup at P08 (15.0 keV) [44] with a detector to sample distance of 560.7 mm (POPC, DPPG, DPPE) or 655.7 mm (DPPC). After deposition, the lipid monolayers were compressed to the target surface tension and kept at this tension. The films were equilibrated for 15 min after reaching the desired tension before acquiring the GIXOS data. To compare the pseudo and the specular reflectivity, GIXOS and conventional XRR measurements were performed on the same samples at OPLS of 12ID-SMI (14.4 keV). For the conventional XRR measurement, the specular reflection was integrated over an area of 0.66×1.00 mm$^2$ (vertical×horizontal) on the detector that was located 1040 mm from the sample. The background was obtained by using the integrated detector signals centered at 1.00 mm to the left and right of the specular position using the same detector area as with the specular condition. In turn, the surface specular reflectivity $R'_{conv}$ [31] was obtained by subtracting the average of the two background signals from the specular signal. The subscript "conv" stands for "conventional XRR". Data reduction of the GIXOS data from both setups were carried out as described in ref. [31], except that here we employed a simplified bulk background subtraction procedure. Instead of using the azimuthally integrated wide angle scattering signal, wide angle GIXOS line cuts at fixed $2\theta = 3.0°$ (P08) and 2.4° (12ID) were used along with the chamber background subtraction described previously. The $Q_{xy}$ positions at $\beta = 0°$, denoted as

*Contact author: chen.shen@desy.de

†Contact author: ocko@bnl.gov

$Q_{xy}|_{\beta=0}$, at these two $2\theta$ angles are 0.4 Å$^{-1}$ and 0.3 Å$^{-1}$, respectively. The background data was fit to an empirical form, $I_{bulk}(|Q|) = y_0 + F\exp(|Q|/t)$, and subtracted from the GIXOS data at the same $|Q|$.

The effective film bending modulus $\kappa$ was obtained by fitting the $Q_{xy}$-dependence of $R^*$ between 0.005 and 0.12 Å$^{-1}$ with the measured surface tension and temperature using Eq. (26) and the roughness factor in Eq. (18), at several $Q_z$ positions simultaneously between 0.1 and 0.6 Å$^{-1}$. Since the film's bending rigidity greatly reduces the thermal roughness on the molecular length scale, the exact choice of the inter-molecular cut-off distance ($a_m$) has no impact on the calculations. For all lipids we set $a_m = 5$ Å, the characteristic distance between two hydrocarbon chains. The intrinsic structure factor $|\Phi(Q_z)|^2$ was computed from $R^*(\beta)$ at various $Q_{xy}|_{\beta=0}$ (P08 data), or at $Q_{xy}|_{\beta=0} = 0.04$ Å$^{-1}$ (12ID data), using Eq. (27) with the obtained bending moduli. For the P08 data, the pseudo reflectivity was calculated using a circular resolution $\delta Q_{xy,R} = 2\times10^{-4}$ Å$^{-1}$ and Eq. (28) and Eq. (30). The corresponding thermal roughness $\sigma_{R,eCWM}$ was calculated using Eq. (23). As discussed in ref. [31], to accurately compare the pseudo reflectivity to the specular reflectivity we must use $R'_{pseudo}$ and $R_{conv}'$ (Eq. 14 in [31]) as both use an identical approach to subtracting the background. For the 12ID comparisons, $R'_{pseudo}$ was calculated using the rectangular resolution settings and the off-specular background settings that match the corresponding specular XRR settings of the reflectometer. The interfacial profiles $\rho_{b,eCWM}(z)$ originating from the eCWM thermal roughness were computed with Eq. (22) and used below. The intrinsic SLD profiles $\rho_{b,0}(z)$ of the films were obtained by fitting the intrinsic structure factors with a simple two-slab density model [45]. This model consists of a hydrocarbon chain slab (subscript "c") and a glycerol-phosphate headgroup slab (subscript "h"), with their respective thickness $D_c$, $D_h$ and respective SLD $\rho_{b,c}$ and $\rho_{b,h}$. The interfaces between the vapor phase and the slab "c", between the slab "c" and "h", and between the slab "h" and the water subphase, are described by three error functions with rms widths $\sigma_{0,air/c}$, $\sigma_{0,c/h}$ and $\sigma_{0,h/w}$, respectively. The subscript 0 represents an "intrinsic width" of the interface without a thermal roughness contribution. Finally, the thermal roughness broadened SLD profiles $\rho_{b,R}(z)$ were computed as $\rho_{b,R}(z) = \rho_{b,\infty}^{-1} \cdot \rho_{b,0}(z) \otimes \rho_{b,eCWM}(z)$, a convolution of the intrinsic SLD profile $\rho_{b,0}(z)$ and the thermal roughness interfacial profile $\rho_{b,eCWM}(z)$.

## IV. Experimental results

Results are presented from X-ray scattering studies on several common phospholipid monolayers at the vapor/liquid interface of aqueous subphases. Their bending moduli and intrinsic structure factors were obtained through the analysis of the surface diffuse X-ray scattering. The pseudo reflectivity is compared with the specular reflectivity measured on the same sample. The validity of the approach is evident from the congruence of the two experimental methods. A representative subset of these results is presented here, with additional data presented in the Supporting Information section S1. In addition to results from monolayer films, water results from 12ID-OPLS are shown in the Supporting Information S2 and complement the published water results from P08 [31]. All the water results follow the CWM description. The Langmuir isotherms of all the films and their isothermal compressibility are presented in Supporting Information S3.

The $Q_{xy}$ dependent diffuse scattering of one lipid monolayer in a fluid phase and another in a crystalline phase are shown in FIG 6 on a log-log scale at four different values of $Q_z$. The data exhibit a visible deviation from the $1/Q_{xy}^{2-\eta}$ dependence conjectured from the CWM roughness factor (black dash-dotted lines) at all $Q_z$. This applies to both the fluid POPC and the crystalline DPPG monolayers. In all cases the scattering falls off faster with $Q_{xy}$ than the assumptions based on the CWM. This deviation from the CWM roughness factor is attributed to a non-zero bending modulus in the roughness factor $\Psi_{eCWM}$ (Eq. (18)). Simultaneously fitting the $Q_{xy}$ dependence at multiple values of $Q_z$ with the eCWM yields $\kappa$ values of (5±3) $k_BT$ for POPC and (15±5) $k_BT$ for DPPG, with the fits shown as solid lines. TABLE IV provides results from the analysis for all films for the parameters; $\kappa$, $L_\kappa$, and the thermal roughness $\sigma_{R,eCWM}$ as calculated for a simulated resolution, $\delta Q_{xy,R} = 2\times10^{-4}$ Å$^{-1}$ (Eq. (23), see section III). Note that $\sigma_{R,eCWM}$ values vary slightly with $Q_z$ through the last term in Eq. (23), with the variation being less than 0.01 Å over a Q$_z$ range up to 1.2 Å$^{-1}$. This variation is negligible compared to the actual value, roughly 3 ~ 4 Å. As shown in TABLE IV, the explicit inclusion of a non-zero bending rigidity in the analysis yields a smaller thermal roughness value by ~ 0.5 Å than the value $\sigma_{R,CWM}$ predicted by the CWM.

*Contact author: chen.shen@desy.de

†Contact author: ocko@bnl.gov

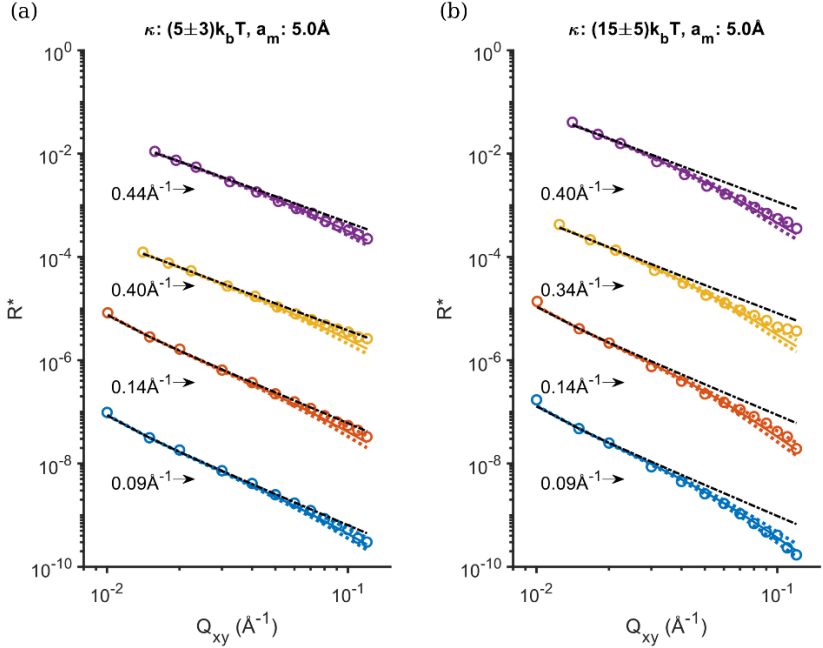

FIG 6. $Q_{xy}$-dependence of the diffuse scattering data from (a) a fluid monolayer (POPC) and (b) a crystalline monolayer (DPPG) at $\gamma = 38$ mN/m (respectively at 20.0 and 22.0°C), and their fits (solid colored lines) by the eCWM using $\kappa$ as the fitting parameter. Data from different $Q_z$ positions are offset by a factor of 10 and $Q_z$ values are indicated with the arrows. The derived $\kappa$ values are $5\pm3k_BT$ for POPC and $15\pm5k_BT$ for DPPG, both shown on figure top. The colored dashed lines on both sides of the solid lines are the dependences that correspond to 68% confidence interval of $\kappa$ as defined by the standard deviation. For comparison, the CWM $Q_{xy}$-dependence ($\kappa = 0$) is shown as black dashed lines.

TABLE IV. The bending modulus $\kappa$, cut-off $L_\kappa$, and the eCWM thermal roughness $\sigma_{R,eCWM}$ calculated for $\delta Q_{xy,R} = 2\times 10^{-4}$ Å$^{-1}$ (also for 12ID data for comparison purpose) of the measured films. The corresponding CWM thermal roughness $\sigma_{R,CWM}$ ($\kappa = 0$) under the same tension and resolution are provided for comparison. $a_m$ is set to 5 Å (see section III).

| film | T [°C] | $\gamma$ [mN/m] | $\kappa$ [$k_BT$] | $L_\kappa$ [Å] | $\sigma_{R,eCWM}$ [Å] | $\sigma_{R,CWM}$ [Å] |
|---|---|---|---|---|---|---|
| POPC | 20.0 | 38.0 | 5±3 | 7.3 | 3.3 | 3.7 |
| DPPC | 22.0 | 38.0 | 10±5 | 10.4 | 3.2 | 3.7 |
| DPPG | 22.0 | 38.0 | 15±5 | 12.7 | 3.2 | 3.7 |
|  |  | 28.0 | 12±5 | 13.2 | 3.7 | 4.3 |
| DPPE | 22.0 | 38.0 | 5±3 | 7.3 | 3.3 | 3.7 |
|  |  | 28.0 | 5±3 | 8.5 | 3.8 | 4.3 |
| DSPC | 20.0 | 28.0 | 20±5 | 17.0 | 3.6 | 4.3 |

The intrinsic structure factors $|\Phi(Q_z)|^2$ of the monolayers are separated from the GIXOS-measured diffuse scattering using the eCWM roughness factor expression (Eq. (27)), after first extracting the values of the bending modulus. FIG 7(a) shows the intrinsic structure factors of a DPPG monolayer as obtained from the GIXOS data at various off-specular positions ($2\theta$, i.e. $Q_{xy}|_{\beta=0}$). Results from additional monolayers are reported in the SI. The inclusion of the eCWM roughness factor yields identical results of $|\Phi(Q_z)|^2$ from the GIXOS data at different $2\theta$-positions, required if the eCWM approach represents the physics correctly.

The intrinsic SLD profile $\rho_{b,0}(z)$ obtained from fitting the $|\Phi(Q_z)|^2$ data are well represented by the classic two slab model (FIG 7(b), red dashed lines) [45,46]. The fitted parameters of the two slab model for all monolayers of this study are compiled into TABLE V. The intrinsic profile $\rho_{b,0}(z)$ exhibits two clearly distinguishable slabs: a high SLD slab for the glycerol-phosphate headgroup (subscript h) and a low SLD hydrocarbon chain slab (subscript c). The rms width of the three interfaces, i.e. between air and the chain slab, between the two slabs, and between the headgroup slab and the subphase are 1.4 Å, 2.8 Å and 1.9 Å, respectively. These rms values are significantly lower than the values of 3 to 3.5 Å [46] that are obtained from fitting the specular reflectivity curves without considering the eCWM roughness contribution. This is because the interface widths in the intrinsic SLD profiles are defined by the distribution of the chemical moieties along the z-direction, while the widths obtained from the specular reflectivity are also broadened by the eCWM roughness.

FIG 7 shows the pseudo reflectivity data along with the intrinsic structure factor data (panel a), and the density profiles obtained from the fits to both data (panel b). The pseudo reflectivity data ($R_{pseudo}/R_F$, FIG 7(a), yellow cross) were derived from the GIXOS diffuse scattering data (SI, Figure S3a) by consideration of the eCWM roughness factor (Eq. (28), Eq. (30)). For calculating $R_{pseudo}$, we choose to use an in-plane resolution $\delta Q_{xy,R}$ of $2\times10^{-4}$ Å$^{-1}$. This is considered a good representation of many liquid scattering instruments, since it corresponds, in the direction of the narrower resolution ($\delta Q_{y,R}$), to a 1 mm high slit at 1 m sample to detector distance calculated at $Q_z \sim 0.8$ Å$^{-1}$ ($\delta Q_{y,R}$ depends on $Q_z$) for an energy of 15 keV. The eCWM roughness factor is reduced to $\exp(-Q_z^2 \sigma_{R,eCWM}^2)$ (FIG 7(a), blue dotted line), with the value of $\sigma_{R,eCWM}$ as tabulated in TABLE IV.

*Contact author: chen.shen@desy.de

†Contact author: ocko@bnl.gov

The profile of $R_{pseudo}/R_F$ (yellow cross, black solid line) falls off faster than $|\Phi(Q_z)|^2$ (FIG 7(a), colored circles, red dash-dotted line), since $R_{pseudo}/R_F$ contains, beside $|\Phi(Q_z)|^2$, the eCWM roughness factor that causes additional fall-off. In real space, the eCWM roughness factor corresponds to the interface profile $\rho_{b,eCWM}(z)$ defined in Eq. (22), an error function with a rms width $\sigma_{R,eCWM}$ (FIG 7(b), blue dotted line). $\rho_{b,eCWM}(z)$ describes the statistically and time averaged width of the interface with the eCWM thermal roughness $\sigma_{R,eCWM}$ over an in-plane length scale that corresponds to $\delta Q_{xy,R}$. The SLD profile $\rho_{b,R}(z)$ that fits the $R_{pseudo}$ (FIG 7(b), black solid line) is a convolution between the intrinsic profile $\rho_{b,0}(z)$ of the film on the surface and $\rho_{b,eCWM}(z)$. It describes the statistically and time averaged film structure on the surface with the eCWM thermal roughness and over the in-plane length scale that corresponds to $\delta Q_{xy,R}$. This SLD profile is smeared from the intrinsic profile due to the thermal roughness (FIG 7(b), black vs red). For a slab model, the interface between two slabs, e.g. m and n, is represented by an error-function shape. The corresponding interfacial width $\sigma_{R,m/n}$ for the SLD profile that fits $R_{pseudo}$ is related to the intrinsic interfacial width $\sigma_{0,m/n}$ as $\sigma_{R,m/n}^2 = \sigma_{0,m/n}^2 + \sigma_{R,eCWM}^2$. The SLD profile obtained by conventional specular reflectivity measurement must be the same $\rho_{b,R}(z)$ since in this work we demonstrate that the pseudo reflectivity curves coincide with the conventional specular reflectivity curves, at equal resolutions.

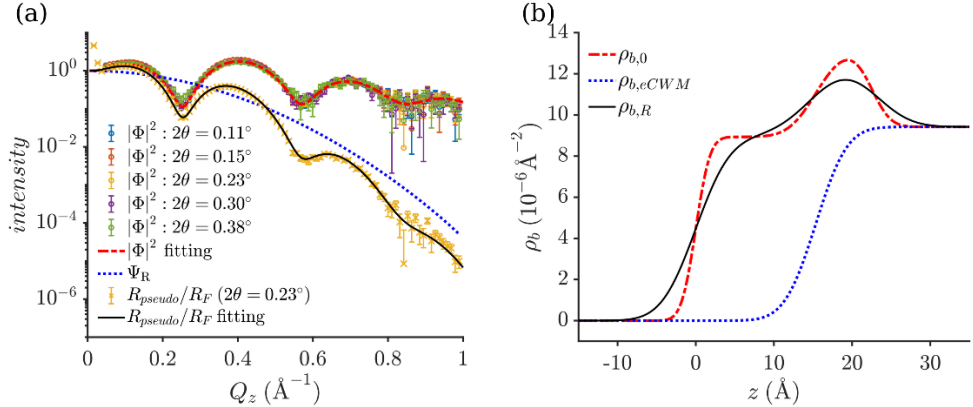

FIG 7. Analysis of the GIXOS data from a DPPG monolayer using the eCWM approach. (a) shows the intrinsic structure factors $|\Phi|^2$ extracted from five $2\theta$-positions using Eq. (27), $R_{pseudo}/R_F$ from $2\theta = 0.23°$ ($Q_{xy}|_{\beta=0} = 0.04$ Å$^{-1}$) for the resolution $\delta Q_{xy,R} = 2\times10^{-4}$ Å$^{-1}$ using Eq. (30), the related fits to them, and the eCWM roughness factor $\Psi_R$ under this resolution. The resolution corresponds to $\sigma_{R,eCWM} = 3.2$ Å. (b) shows the intrinsic SLD profile $\rho_{b,0}(z)$ that fits $|\Phi|^2$, the thermal roughness profile $\rho_{b,eCWM}(z)$ for this roughness factor, and the SLD profile $\rho_{b,R}(z)$ that corresponds to $R_{pseudo}/R_F$.

*Contact author: chen.shen@desy.de

†Contact author: ocko@bnl.gov

TABLE V. Parameters of the two-slab model that describes the intrinsic SLD profile of the measured monolayers [45], as obtained from fitting their intrinsic structure factors. See the definition of the abbreviations in the experimental details.

| film | T [°C] | γ [mN/m] | $\sigma_{0,air/c}$ [Å] | $D_c$ [Å] | $\rho_{b,c}$ [$10^{-6}$ Å$^{-2}$] | $\sigma_{0,c/h}$ [Å] | $D_h$ [Å] | $\rho_{b,h}$ [$10^{-6}$ Å$^{-2}$] | $\sigma_{0,h/w}$ [Å] |
|---|---|---|---|---|---|---|---|---|---|
| POPC | 20.0 | 38.0 | 1.9±0.1 | 12.7±0.1 | 8.3±0.1 | 2.8±0.1 | 7.3±0.1 | 11.6±0.1 | 1.6±0.1 |
| DPPC | 22.0 | 38.0 | 1.7±0.1 | 14.6±0.1 | 9.0±0.1 | 2.3±0.1 | 8.3±0.1 | 12.3±0.1 | 1.0±0.1 |
| DPPG | 22.0 | 38.0 | 1.4±0.1 | 15.4±0.1 | 8.9±0.1 | 2.8±0.1 | 7.1±0.1 | 13.2±0.1 | 1.9±0.1 |
| DPPG | 22.0 | 28.0 | 1.7±0.1 | 16.0±0.1 | 9.0±0.1 | 2.8±0.1 | 7.1±0.1 | 12.9±0.1 | 1.8±0.1 |
| DPPE | 22.0 | 38.0 | 1.2±0.3 | 16.3±0.2 | 9.1±0.1 | 2.3±0.7 | 9.1±0.5 | 12.9±0.4 | 1±1 |
| DPPE | 22.0 | 28.0 | 1.6±0.1 | 17.3±0.1 | 9.4±0.1 | 1.7±0.1 | 8.6±0.1 | 12.8±0.1 | 1.1±0.1 |
| DSPC | 20.0 | 28.0 | 1.8±0.1 | 17.9±0.1 | 9.3±0.3 | 1.8±0.2 | 8.6±0.1 | 13.2±0.3 | 1.0±0.6 |

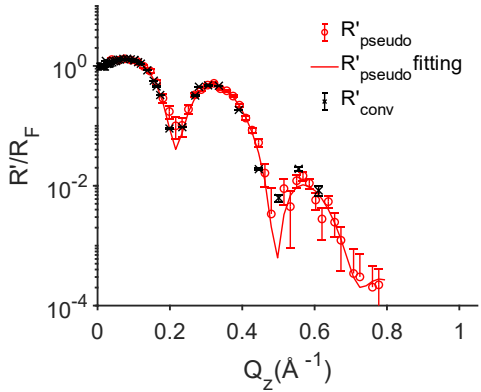

FIG 8. Pseudo reflectivity $R'_{pseudo}$ and conventional specular reflectivity $R'_{conv}$ from the very same DSPC monolayer at $\gamma$ = 28 mN/m (20°C). $R'_{pseudo}$ is derived for the same detector slit settings and the background subtraction of the specular reflectometry experiment (see section III).

In the following, experimental verification is presented in order to demonstrate that these two reflectivity curves overlap when the bending rigidity contribution to the capillary waves is considered in the roughness factor. FIG 8 shows $R'_{pseudo}$ and $R'_{conv}$ measured from the same DSPC monolayer. The background subtraction in the conventional specular reflectivity demands $R'$ to be used instead of $R$ (see section III). The two curves overlap in their shared $Q_z$-range, and $R'_{pseudo}$ covers a larger $Q_z$-range [31]. Note that $R'_{pseudo}$ here is derived using the integral expression of $\Psi_{eCWM}$ (Eq. (18)) for the rectangular slit settings that match the slit settings in the conventional specular reflectivity measurements. Our findings now remove the apparent deviations that were reported previously in pioneering studies on the comparison between GIXOS analysis and specular XRR [21,33,47]. The deviations were likely due to a combination of factors, that include (1) specifics of the background subtraction algorithm, (2) omitting the roughness factor in the analysis [21,33], (3) neglecting the bending rigidity [42] and (4) the specifics of the description of the headgroup density profile, namely by using a single Gaussian [42] rather than a headgroup box of constant density [47].

## V. Discussion

In this section we provide a perspective on the principal findings of this study. These include the application of GIXOS analysis on Langmuir monolayers to obtain the bending rigidity, intrinsic film structure and pseudo reflectivity, and the limit of the theory at large $Q_z$ when the parameter $\eta$ exceeds a value of 2.

The analysis presented demonstrates the importance of considering the effect of the bending rigidity on thermal roughness and how this modifies the interpretation of specular reflectivity data for film characterization. The bending rigidity changes the capillary wave roughness through the cut-off $Q_\kappa$, and this cut-off can only be determined through the diffuse scattering results [20,25,26,28,29] and not from the specular reflectivity. Applying the GIXOS method allows one to obtain the bending rigidity from the $Q_{xy}$ dependence of the diffuse scattering. This straightforwardly delivers the intrinsic structure factor, that does not contain the thermal roughness contribution. In FIG 7(b) the analyzed intrinsic density profile of the film was plotted along with the density profile that fits the reflectivity. The structural features are more apparent after the effects of the thermal roughness broadening are removed. Note that similar results were reported for hydrated fluid lipid multilayers [48,49]. Another benefit of the method is the higher spatial resolution as compared to conventional XRR since the results extend up to a larger $Q_z$ : conventional specular reflectometry typically provides reflectivity to ~ 0.7 Å$^{-1}$, while with GIXOS – pseudo XRR the range is often extended to beyond 1.0 Å$^{-1}$. This extended $Q_z$-range allows more parameters to be determined through the fitting. For instance, with the GIXOS – pseudo XRR approach it is possible to obtain up to seven free fitting parameters using a two-slab model (3 rms widths, 2 densities, and 2 thicknesses). In contrast, conventional XRR fitting demands some of these parameters to be constrained if small uncertainties for the parameters are required. A typical constraint is to set the three widths to be equal [45,46] (essentially, fitting only one of them). The extended $Q_z$-range is to be attributed to two principal reasons. Firstly, with conventional reflectometry, the strong bulk scattering background hampers the extraction of the weak reflectivity signal ($<10^{-9}$) in the extended $Q_z$-range. In contrast, with GIXOS the bulk background scattering is relatively small due to the low penetration depth provided by the grazing incidence condition (~ 10 nm at $\alpha = 0.85\alpha_c$) [31]. Secondly, at these extended $Q_z$ ranges the surface thermal roughness broadens the transverse scattering ($Q_{xy}$) peak around the specular position due to the $Q_z$ dependence of $\eta$, and this makes the specular signal weaker [31]. Using the GIXOS method permits one to capture the weak surface diffuse scattering at larger $Q_z$ by integrating the signal over a large solid-angle and this way enhances the signal over the background.

Obtaining the value of the bending rigidity for several commonly used phospholipid monolayers is among the most significant benefits from the type of study presented here. Data are compiled into TABLE IV. The value of 5 $k_BT$ for the fluid POPC monolayer compares well to the values reported from studies on fluid PC bilayers of 15~20 $k_BT$ (note that a bilayer consists of two monolayers, resulting in an approximately factor of two for the stiffness of a bilayer) [2,9,20,50]. The value for the crystalline phase monolayer was found to be 10 ~ 20 $k_BT$, similar to the result from an earlier diffuse scattering measurement on a DPPC monolayer on the water surface [51]. This is close to the estimated value from the Langmuir isotherm using the expression, $\kappa = \frac{\varepsilon D_c^2}{12}$, assuming a uniform lateral pressure profile over the chain slab [50], e.g. 12 $k_BT$ for DPPC at 35 mN/m. The thickness of the chain slab was obtained from the fitting (TABLE V), while the value for the compression moduli $\varepsilon$ were found to be between 0.2 and 0.3 N/m, as extracted from the slope of the monolayer isotherm, respectively at $\gamma = 38$ and 28 mN/m (Supporting Information S3). Note that our experimental value is an order of magnitude lower than the published data on the gel phase bilayer (~300$k_BT$ [9,20]). The high bilayer value is possibly related to the coupling between the two monolayers – sometimes referred as leaflets – in a bilayer [50]. The present studies cannot resolve this discrepancy and it will require further study.

We noticed a limitation in the mathematical formalism of all existing capillary wave models at large $Q_z$, since GIXOS allowed us to observe the continuously developing surface scattering signal into a large $Q_z$ range that was not achievable with conventional XRR. The problem shows up because the mathematical description of the scattering cross section becomes unphysical when the dimensionless parameter, $\eta = \frac{k_BT}{2\pi\gamma}Q_z^2$, exceeds a value of 2. The specular reflection ($Q_{xy} = 0$) is properly described by the cusp-like algebraic singularity of $\chi(\eta)Q_{xy}^{\eta-2}$ of the CWM term of the cross section, as long as $\eta < 2$ [14,15,18]. Beyond that ($\eta \geq 2$) the model becomes unphysical since it would predict that the specular reflection is no longer stronger than the scattering at off-specular positions. Moreover, it is also unphysical that $\chi(\eta)$ is not continuous across $\eta = 2$ due to the behavior of $\Gamma\left(1 - \frac{\eta}{2}\right)$. This condition defines a maximal $Q_z$ value up to which the surface scattering can be reasonably described by the existing capillary wave models, including the eCWM. This $Q_z$ limit decreases with reducing tension (FIG 9). Of note, our measurements at 28 mN/m (FIG 10) and at even lower tension (data not shown) provide the evidences that the surface scattering signal is continuous also for $\eta \geq 2$. Since the mathematical model fails for the $Q_z$ range where $\eta \geq 2$ the data analysis cannot be performed above this range. This constraint by $\eta$ on the model requires further elaboration.

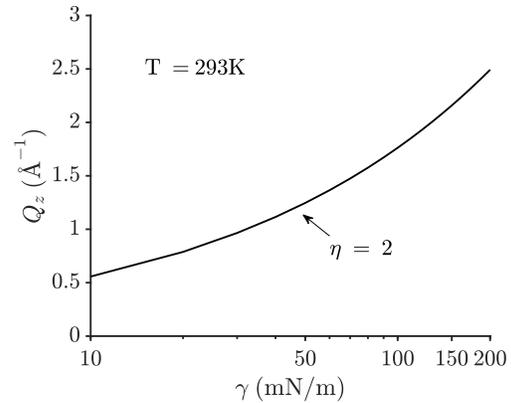

FIG 9. $Q_z$-value at $\eta = 2$ as a function of surface tension at 293 K.

*Contact author: chen.shen@desy.de

†Contact author: ocko@bnl.gov

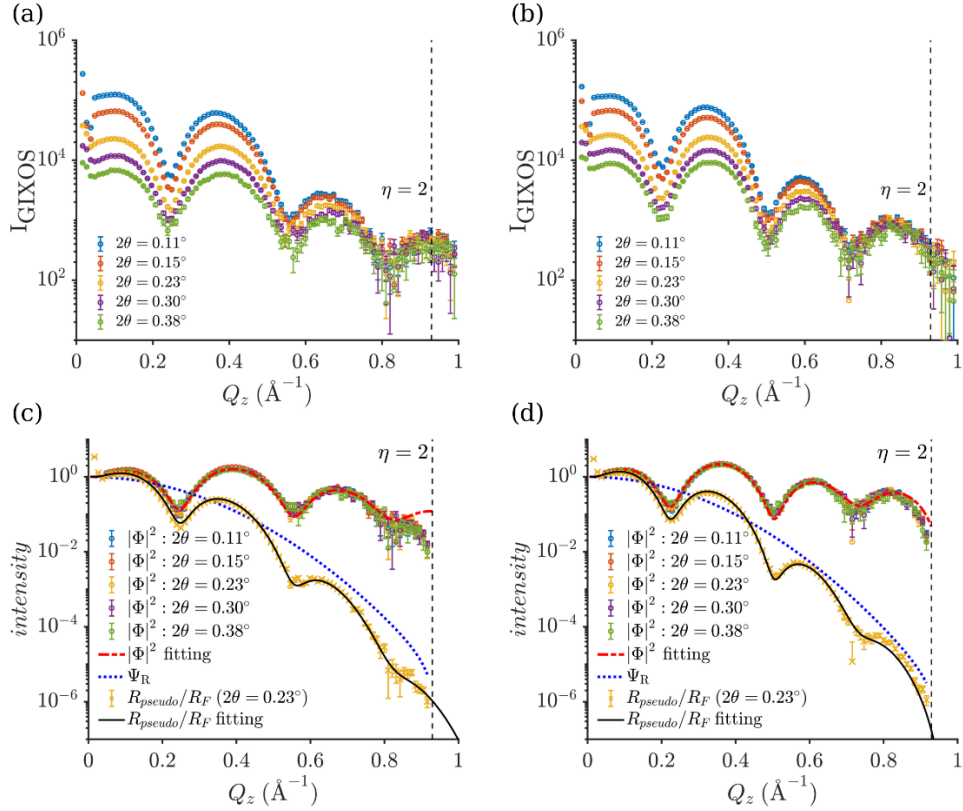

FIG 10. GIXOS data obtained from different $2\theta$ positions from (a) DPPG and (b) DPPE monolayers at $\gamma = 28$ mN/m and their analysis using the eCWM (c, d). The analysis can only be performed until $Q_z = 0.93$ Å$^{-1}$ where $\eta = 2$ at 28 mN/m (vertical dashed line). The analysis plot (c, d) shows the intrinsic structure factors $|\Phi|^2$ extracted from these GIXOS data, the pseudo reflectivity $R_{pseudo}/R_F$ and the fits to them, and the eCWM roughness factor $\Psi_R$. Values for $R_{pseudo}/R_F$ and $\Psi_R$ are calculated for the resolution $\delta Q_{xy,R} = 2 \times 10^{-4}$ Å$^{-1}$, and their corresponding $\sigma_{R,eCWM}$ values are compiled into TABLE IV.

Finally, we will note the normalization of the data. In our previous study [31], the normalization factor is correctly obtained from the direct beam intensity. In the present study a discrepancy was found for the case of crystalline monolayer films. The intrinsic structure factor in Eq. (27) should also approach unity at small $Q_z$ to ensure the reflectivity approaches unity at the critical angle. This works out well for all non-crystalline samples such as POPC. However, a discrepancy showed up for crystalline films (DPPC, DPPG, DPPE). Normalizing their GIXOS data with direct intensity resulted in an under-scaled structure factor by up to 40% and this would require further studies.

## V. CONCLUSIONS

The paper presents an extension of the standard Capillary Wave Model (CWM) that explicitly includes the effect of a non-zero bending modulus at a liquid interface. The extended CWM (eCWM) provides a unified analytical form of the thermal roughness factor, to be used to describe the specular X-ray reflectivity and the off-specular diffuse scattering. This single form is valid for both zero and finite bending rigidity cases, and recovers the standard CWM expression [18] at zero bending rigidity. With this approach, the fitting of the bending modulus of an interfacial film is now possible from the analysis of GIXOS data, and subsequently it provides the surface normal intrinsic structure factor beyond 1.0 Å$^{-1}$. The latter results, combined with fitting analysis, is used to obtain the surface normal structure more accurately than with conventional XRR.

## ACKNOWLEDGMENTS
We acknowledge DESY (Hamburg, Germany), a member of the Helmholtz Association HGF, for the provision of experimental facilities. Parts of this research were carried out at PETRA III and we thank Dr. F. Bertram and Mr. R. Kirchhof for assistance in using P08, Dr. M. Lippmann and Mrs. M. Dahdouli in using the chemistry lab. Beamtime was allocated for proposal I-20221110. Eiger2 detector from AG. Magnussen (Kiel University) is funded through ErUM-Pro 05K19FK2 (Murphy) by German Federal Ministry of Education and Research. Parts of this research used the Open Platform Liquid Surface (OPLS) endstation of the Soft Matter Interfaces beamline (12ID-SMI) of the National Synchrotron Light Source II, a US Department of Energy (DOE) Office of Science user facility operated for the DOE Office of Science by Brookhaven National Laboratory (BNL) under contract No. DE-SC0012704. At NSLS II we gratefully acknowledge Dr. D. Bacescu for the design of the OPLS instrument and Mr. R. Greene, Mr. B. Marino, Mr. S. Meyer and Mr. Z. Yin for technical support. We thank Dr. M. Fukuto from BNL, for discussions on the theory and a critical reading of the manuscript. B.K. thanks PhyLife@SDU and the Danish Agency for Science, Technology and Innovation for funding the instrument center DanScatt. We thank Dr. L. Pithan, Mrs. T. Jakkampudi, Dr. B. Bharatiya from DESY, Mr. A. Wright from Stony Brook University for assisting in the measurements, Prof. J. F. Nagle and Prof. S. Tristram-Nagle from Carnegie Mellon University for discussion. Finally, C.S. thanks H. Matjes and G. Matjes, W. Spargel, and TIDE Treibholz und Feinkost, Planten un Blomen in Hamburg for providing inspiration.

*Contact author: chen.shen@desy.de

†Contact author: ocko@bnl.gov

## APPENDIX A: Derivation from power spectral density to height-height correlation function

The height-height correlation function of a rough surface is a 2D inverse Fourier transform of the PSD of the surface:

$$\frac{1}{(2\pi)^2} \int_{Q_{min}}^{Q_{max}} d\boldsymbol{Q}_{xy} \exp(-i\boldsymbol{Q}_{xy} \cdot \boldsymbol{r}_{xy}) \langle \tilde{h}(\boldsymbol{Q}_{xy})\tilde{h}(-\boldsymbol{Q}_{xy}) \rangle$$

$$= \frac{1}{(2\pi)^2} \int_{Q_{min}}^{Q_{max}} d\boldsymbol{Q}_{xy} \exp(-i\boldsymbol{Q}_{xy} \cdot \boldsymbol{r}_{xy}) \frac{1}{A} \int_{r_{min}}^{r_{max}} d\boldsymbol{r}''_{xy} \exp(i\boldsymbol{Q}_{xy} \cdot \boldsymbol{r}''_{xy}) \langle h(\boldsymbol{r}''_{xy})h(\boldsymbol{0}) \rangle$$

$$= \frac{1}{A(2\pi)^2} \int_{r_{min}}^{r_{max}} d\boldsymbol{r}''_{xy} \langle h(\boldsymbol{r}''_{xy})h(\boldsymbol{0}) \rangle \int_{Q_{min}}^{Q_{max}} d\boldsymbol{Q}_{xy} \exp(i\boldsymbol{Q}_{xy} \cdot [\boldsymbol{r}''_{xy} - \boldsymbol{r}_{xy}])$$

$$= \frac{1}{A(2\pi)^2} \int_{r_{min}}^{r_{max}} d\boldsymbol{r}''_{xy} \langle h(\boldsymbol{r}''_{xy})h(\boldsymbol{0}) \rangle \cdot (2\pi)^2 \delta(\boldsymbol{r}''_{xy} - \boldsymbol{r}_{xy}) = \frac{1}{A} \langle h(\boldsymbol{r}_{xy})h(\boldsymbol{0}) \rangle$$

For the PSD $\langle \tilde{h}(\boldsymbol{Q}_{xy})\tilde{h}(-\boldsymbol{Q}_{xy}) \rangle = \frac{1}{A} \cdot \frac{k_B T}{\Delta\rho_m g + \gamma Q_{xy}^2 + \kappa Q_{xy}^4}$, the height-height correlation function is

$$\langle h(\boldsymbol{r}_{xy})h(\boldsymbol{0}) \rangle = \frac{A}{(2\pi)^2} \int_0^{Q_{max}} d\boldsymbol{Q}_{xy} \exp(-i\boldsymbol{Q}_{xy} \cdot \boldsymbol{r}_{xy}) \left[ \frac{1}{A} \cdot \frac{k_B T}{\Delta\rho_m g + \gamma Q_{xy}^2 + \kappa Q_{xy}^4} \right]$$

$$= \frac{1}{(2\pi)^2} \int_0^{Q_{max}} \frac{k_B T}{\Delta\rho_m g + \gamma Q_{xy}^2 + \kappa Q_{xy}^4} \cdot Q_{xy} dQ_{xy} \cdot \int_0^{2\pi} \exp(-iQ_{xy} r_{xy} \cos\theta) d\theta$$

$$= \frac{k_B T}{2\pi} \int_0^{Q_{max}} \frac{J_0(Q_{xy} r_{xy})}{\Delta\rho_m g + \gamma Q_{xy}^2 + \kappa Q_{xy}^4} Q_{xy} dQ_{xy}$$

where $d\boldsymbol{Q}_{xy} = Q_{xy} dQ_{xy} d\theta$ is applied to convert the vector integration into scalar integration. $\theta$ is the angle between $\boldsymbol{Q}_{xy}$ and $\boldsymbol{r}_{xy}$. The integration definition $J_0(x) = J_0(-x) = \frac{1}{\pi} \int_0^{\pi} \exp(ix \cos\theta) d\theta$ is applied [52].

The upper limit of the integral $Q_{max}$ is substituted by infinity when $r_{xy}$ is replaced by $r'_{xy} = \sqrt{r_{xy}^2 + Q_{max}^{-2}}$ [35]:

$$\langle h(\boldsymbol{r}'_{xy})h(\boldsymbol{0}) \rangle = \frac{k_B T}{2\pi} \int_0^{\infty} \frac{J_0(Q_{xy} r'_{xy})}{\Delta\rho_m g + \gamma Q_{xy}^2 + \kappa Q_{xy}^4} Q_{xy} dQ_{xy}$$

McClain, et. al. solved the correlation function above as the following [34], by applying $\int_0^{\infty} \frac{x^{\nu+1} J_\nu(ax)}{x^2+b^2} dx = b^\nu K_\nu(ab)$ [53], where $J_\nu$ and $K_\nu$ are the $\nu$-th order of the Bessel function of the 1st kind and of the modified Bessel function of the 2nd kind, respectively.

*Contact author: chen.shen@desy.de

†Contact author: ocko@bnl.gov

$$\langle h(\boldsymbol{r}'_{xy})h(\boldsymbol{0})\rangle = \frac{k_B T}{2\pi}\int_0^\infty \frac{J_0(Q_{xy}r'_{xy})Q_{xy}}{\Delta\rho_m g + \gamma Q_{xy}^2 + \kappa Q_{xy}^4}dQ_{xy}$$

$$= \frac{k_B T}{2\pi}\int_0^\infty \frac{J_0(Q_{xy}r'_{xy})Q_{xy}}{\sqrt{\gamma^2 - 4\Delta\rho_m g\kappa}}\cdot\left[\frac{1}{Q_{xy}^2 + \frac{\gamma - \sqrt{\gamma^2 - 4\Delta\rho_m g\kappa}}{2\kappa}} - \frac{1}{Q_{xy}^2 + \frac{\gamma + \sqrt{\gamma^2 - 4\Delta\rho_m g\kappa}}{2\kappa}}\right]dQ_{xy}$$

$$= \frac{k_B T}{2\pi\sqrt{\gamma^2 - 4\Delta\rho_m g\kappa}}\cdot\left[K_0\left(r'_{xy}\sqrt{\frac{\gamma - \sqrt{\gamma^2 - 4\Delta\rho_m g\kappa}}{2\kappa}}\right) - K_0\left(r'_{xy}\sqrt{\frac{\gamma + \sqrt{\gamma^2 - 4\Delta\rho_m g\kappa}}{2\kappa}}\right)\right]$$

Note the rigidity related cut-off $L_\kappa = \sqrt{\kappa/\gamma}$ and the gravitational cut-off $L_g = \sqrt{\gamma/\Delta\rho_m g}$. Introducing $s = \sqrt{1 - 4\Delta\rho_m g\kappa/\gamma^2} = \sqrt{1 - 4(L_\kappa/L_g)^2}$ yields:

$$\langle h(\boldsymbol{r}'_{xy})h(\boldsymbol{0})\rangle = \frac{k_B T}{2\pi\gamma\cdot s}\cdot\left[K_0\left(\frac{r'_{xy}}{L_\kappa}\sqrt{\frac{1-s}{2}}\right) - K_0\left(\frac{r'_{xy}}{L_\kappa}\sqrt{\frac{1+s}{2}}\right)\right]$$

For a typical soft matter thin film, $\gamma$ is less than 100 mN/m. $\kappa$ is typically less than $1000k_B T$. Therefore $L_\kappa$ is in the order between some to some hundreds of nanometers. $L_g$ is in the order of some millimeters. Therefore $\frac{\Delta\rho_m g}{\gamma} \ll \frac{\gamma}{4\kappa}$, and $\frac{L_\kappa}{L_g} \to 0$. Accordingly, $s = \sqrt{1 - 4(L_\kappa/L_g)^2} \approx \sqrt{1 - 4\left(\frac{L_\kappa}{L_g}\right)^2 + \left[2\left(\frac{L_\kappa}{L_g}\right)^2\right]^2} = 1 - 2\left(\frac{L_\kappa}{L_g}\right)^2$, $\sqrt{\frac{1-s}{2}} = \frac{L_\kappa}{L_g}$ and $\sqrt{\frac{1+s}{2}} \approx 1$. The height-height correlation function is simplified to $\langle h(\boldsymbol{r}'_{xy})h(\boldsymbol{0})\rangle = \frac{k_B T}{2\pi\gamma}\left[K_0\left(\frac{r'_{xy}}{L_g}\right) - K_0\left(\frac{r'_{xy}}{L_\kappa}\right)\right]$.

## APPENDIX B: Statistical average of the surface topology in differential scattering cross section

The differential cross section is the squared modulus of the Fourier transform on the SLD distribution in the illuminated volume:

$$\frac{d\sigma}{d\Omega} = \left(\frac{\rho_{b,\infty}}{Q_z}\right)^2\cdot|\Phi(Q_z)|^2\cdot\iint \exp\left(iQ_z[h(\boldsymbol{r}_{xy}) - h(\boldsymbol{r}''_{xy})] + i\boldsymbol{Q}_{xy}\cdot(\boldsymbol{r}_{xy} - \boldsymbol{r}''_{xy})\right)d\boldsymbol{r}_{xy}d\boldsymbol{r}''_{xy}\cdot|t(\alpha)|^2|t(\beta)|^2$$

Note that $\boldsymbol{r}''_{xy}$ and $\boldsymbol{r}_{xy}$ are two positions in the surface plane. The time averaged surface height displacement is isotropic in the in-plane direction. $h(\boldsymbol{r}_{xy}) - h(\boldsymbol{r}''_{xy})$ only depends on the relative distance $\boldsymbol{r}_{xy} - \boldsymbol{r}''_{xy}$. Therefore the integrals over $\boldsymbol{r}''_{xy}$ and over $\boldsymbol{r}_{xy}$ can be separately performed. Note the use of the time and ensemble average $\langle\exp(iQ_z[h(\boldsymbol{r}_{xy}) - h(\boldsymbol{0})])\rangle$:

$$\frac{d\sigma}{d\Omega} = \left(\frac{\rho_{b,\infty}}{Q_z}\right)^2\cdot|\Phi(Q_z)|^2\cdot\int\langle\exp(iQ_z[h(\boldsymbol{r}_{xy}) - h(\boldsymbol{0})])\rangle\exp(i\boldsymbol{Q}_{xy}\cdot\boldsymbol{r}_{xy})d\boldsymbol{r}_{xy}\int_A d\boldsymbol{r}''_{xy}\cdot|t(\alpha)|^2|t(\beta)|^2$$

$$= \frac{A_0}{\sin\alpha}\left(\frac{\rho_{b,\infty}}{Q_z}\right)^2\cdot|\Phi(Q_z)|^2\cdot\int\langle\exp(iQ_z[h(\boldsymbol{r}_{xy}) - h(\boldsymbol{0})])\rangle\exp(i\boldsymbol{Q}_{xy}\cdot\boldsymbol{r}_{xy})d\boldsymbol{r}_{xy}\cdot|t(\alpha)|^2|t(\beta)|^2$$

$A$ is the unit illuminated film area. Here we apply $\int_A d\boldsymbol{r}''_{xy} = A_0/\sin\alpha$ where $A_0$ and $\alpha$ are the cross section area of the incident beam and the incident angle, respectively. The statistical average on the exponential term has the property of $\langle\exp(ix)\rangle = \exp\left(-\frac{1}{2}\langle x^2\rangle\right)$. This yields:

*Contact author: chen.shen@desy.de

†Contact author: ocko@bnl.gov

$$\frac{d\sigma}{d\Omega} = \frac{A_0}{\sin\alpha}\left(\frac{\rho_{b,\infty}}{Q_z}\right)^2 \cdot |\Phi(Q_z)|^2 \cdot \int \exp\left(-\frac{1}{2}Q_z^2 \langle[h(\boldsymbol{r}_{xy}) - h(\boldsymbol{0})]^2\rangle\right) \exp(i\boldsymbol{Q}_{xy}\cdot\boldsymbol{r}_{xy})\,d\boldsymbol{r}_{xy} \cdot |t(\alpha)|^2|t(\beta)|^2$$

Applying $\langle[h(\boldsymbol{r}_{xy}) - h(\boldsymbol{0})]^2\rangle = 2\langle[h(\boldsymbol{r}_{xy})]^2\rangle - 2\langle h(\boldsymbol{r}_{xy})h(\boldsymbol{0})\rangle$ gives:

$$\frac{d\sigma}{d\Omega} = A_0\left(\frac{Q_c}{2Q_z}\right)^4 |t_\alpha|^2|t_\beta|^2 \cdot |\Phi(Q_z)|^2 \cdot \frac{Q_z^2}{16\pi^2 \sin\alpha} \exp(-Q_z^2\langle h^2\rangle) \int \exp(Q_z^2 \langle h(\boldsymbol{r}_{xy})h(\boldsymbol{0})\rangle) \exp(i\boldsymbol{Q}_{xy}\cdot\boldsymbol{r}_{xy})\,d\boldsymbol{r}_{xy}$$

Here $\frac{Q_z^2}{16\pi^2 \sin\alpha}$ is introduced such that the differential roughness factor is dimensionless.

### APPENDIX C: Derivation of the eCWM differential roughness factor

The expression of the differential roughness factor $\psi_{eCWM}(Q_{xy}, Q_z)$ for the eCWM is obtained by inserting $\langle h(\boldsymbol{r}_{xy})h(\boldsymbol{0})\rangle$ and $\langle h^2\rangle$ from Eq. (8) and Eq. (3) into Eq. (7):

$$\psi_{eCWM}(Q_{xy}, Q_z)$$
$$= \frac{Q_z^2}{16\pi^2 \sin\alpha} \cdot \left(\frac{1}{Q_{max}}\right)^\eta \cdot \exp\left[\eta K_0\left(\frac{1}{L_\kappa Q_{max}}\right)\right] \cdot \int_0^\infty (r'_{xy})^{-\eta} \exp\left[-\eta K_0\left(\frac{r'_{xy}}{L_\kappa}\right)\right] \exp(i\boldsymbol{Q}_{xy}\cdot\boldsymbol{r}'_{xy})\,d\boldsymbol{r}_{xy}$$
$$= \frac{Q_z^2}{16\pi^2 \sin\alpha} \cdot \left(\frac{1}{Q_{max}}\right)^\eta \cdot \exp\left[\eta K_0\left(\frac{1}{L_\kappa Q_{max}}\right)\right] \cdot \int_0^\infty\int_0^{2\pi} (r'_{xy})^{-\eta} \exp\left[-\eta K_0\left(\frac{r'_{xy}}{L_\kappa}\right)\right] \cdot \exp(iQ_{xy} r'_{xy} \cos\theta)\,r'_{xy}\,dr_{xy}\,d\theta$$
$$= \frac{Q_z^2}{16\pi^2 \sin\alpha} \cdot \left(\frac{1}{Q_{max}}\right)^\eta \cdot \exp\left[\eta K_0\left(\frac{1}{L_\kappa Q_{max}}\right)\right] \cdot 2\pi \int_0^\infty (r'_{xy})^{1-\eta} \exp\left[-\eta K_0\left(\frac{r'_{xy}}{L_\kappa}\right)\right] J_0(Q_{xy} r'_{xy})\,dr_{xy}$$

Here three equalities are applied: (a) the integration on the vector $\int(\ldots) \exp(i\boldsymbol{Q}_{xy}\cdot\boldsymbol{r}'_{xy})\,d\boldsymbol{r}_{xy} = \int\int_0^{2\pi}(\ldots)\exp(iQ_{xy} r'_{xy} \cos\theta)\,r'_{xy}\,dr_{xy}\,d\theta$, (b) $J_0(x) = J_0(-x) = \frac{1}{\pi}\int_0^\pi \exp(ix\cos\theta)\,d\theta$ [52], and (c) the relation $\int_\pi^{2\pi} \exp(ix\cos\theta)\,d\theta = \int_0^\pi \exp(-ix\cos\theta)\,d\theta$. The in-plane distance is replaced by the molecular size-adjusted distance $r'_{xy}$ such that lower limit of the integral is zero [35].

This integral cannot be computed numerically or analytically since the integrand oscillates with an increasing amplitude for large $Q_{xy} r'_{xy}$. In order to numerically compute the integration, we separate out the CWM's integration $\int_0^\infty (r'_{xy})^{1-\eta} J_0(Q_{xy} r'_{xy})\,dr_{xy} = \frac{2^{1-\eta}\cdot\Gamma\left(1-\frac{\eta}{2}\right)}{\Gamma\left(\frac{\eta}{2}\right)} \cdot \frac{1}{Q_{xy}^{2-\eta}}$ [11,54] from the integrand and rearrange the differential roughness factor:

$$\psi_{eCWM}(Q_{xy}, Q_z)$$
$$= \frac{Q_z^2}{16\pi^2 \sin\alpha} \cdot \left(\frac{1}{Q_{max}}\right)^\eta \cdot \exp\left[\eta K_0\left(\frac{1}{L_\kappa Q_{max}}\right)\right]$$
$$\cdot \left\{2\pi \int_0^\infty (r'_{xy})^{1-\eta} J_0(Q_{xy} r'_{xy})\,dr_{xy} + 2\pi \int_0^\infty (r'_{xy})^{1-\eta} \left[\exp\left(-\eta K_0\left(\frac{r'_{xy}}{L_\kappa}\right)\right) - 1\right] J_0(Q_{xy} r'_{xy})\,dr_{xy}\right\}$$
$$\approx \frac{Q_z^4}{16\pi^2 \sin\alpha} \cdot \left(\frac{1}{Q_{max}}\right)^\eta \cdot \exp\left[\eta K_0\left(\frac{1}{L_\kappa Q_{max}}\right)\right]$$
$$\cdot \left\{\frac{2\pi}{Q_z^2} \cdot \frac{2^{1-\eta}\cdot\Gamma\left(1-\frac{\eta}{2}\right)}{\Gamma\left(\frac{\eta}{2}\right)} \cdot \frac{1}{Q_{xy}^{2-\eta}} + \frac{2\pi}{Q_z^2}\int_0^{8L_\kappa} (r'_{xy})^{1-\eta} \left[\exp\left(-\eta K_0\left(\frac{r'_{xy}}{L_\kappa}\right)\right) - 1\right] J_0(Q_{xy} r'_{xy})\,dr_{xy}\right\}$$

*Contact author: chen.shen@desy.de

†Contact author: ocko@bnl.gov

$$= \frac{Q_z^4}{16\pi^2 \sin \alpha} \cdot \left(\frac{1}{Q_{max}}\right)^\eta \cdot \exp\left[\eta K_0\left(\frac{1}{L_\kappa Q_{max}}\right)\right] \cdot \left[\frac{\chi(\eta)}{Q_{xy}^{2-\eta}} + \frac{C'(Q_{xy}, \eta, L_\kappa)}{Q_z^2}\right]$$

Here the limited integration range to $8L_\kappa$ is sufficiently accurate to get a good approximation since the integrand quickly approaches zero when the in-plane distance increases towards $8L_\kappa$. This property of the integrand is described in detail in APPENDIX D

## APPENDIX D: Properties of $C'(Q_{xy}, \eta, L_\kappa)$ and $C(\eta, L_\kappa)$

In this section the properties of $C'(Q_{xy}, \eta, L_\kappa)$ and its integration $C(\eta, L_\kappa)$ around the specular condition under circular resolution are analyzed to evaluate their contribution.

The term $C'(Q_{xy}, \eta, L_\kappa)$ in the differential scattering cross section is

$$C'(Q_{xy}, \eta, L_\kappa) = 2\pi \int_0^\infty (r'_{xy})^{1-\eta} \left[\exp\left(-\eta K_0\left(\frac{r'_{xy}}{L_\kappa}\right)\right) - 1\right] J_0(Q_{xy} r'_{xy}) dr'_{xy}$$

The integrand consists of an exponent related to $r'_{xy}$, $J_0(Q_{xy} r'_{xy})$, and $(r'_{xy})^{1-\eta}$ that increases less than a linear dependence on $r'_{xy}$. $J_0(Q_{xy} r'_{xy}) = 1$ when $Q_{xy} r'_{xy} \to 0$, and it oscillates around zero with a decreasing amplitude when $Q_{xy} r'_{xy}$ increases [55]. $K_0(x)$ in the exponent approaches positive infinity when the distance approaches 0, and monotonically decreases to zero at a larger distance. Therefore $(r'_{xy})^{1-\eta} \left[\exp\left(-\eta K_0\left(\frac{r'_{xy}}{L_\kappa}\right)\right) - 1\right]$ quickly approaches zero with the increasing distance. The integration range in $r'_{xy}$ hence can be limited to a distance where the integrand is almost zero for the $Q_{xy}$ that is relevant for the X-ray measurements.

In the following the relevant range of $r'_{xy}$ is determined. This is related to $J_0(x)$ and hence depends on the off specular angle $Q_{xy}$. Increasing $Q_{xy}$ reduces the oscillation period of $J_0(x)$ in term of $r'_{xy}$ and causes a significant modulation in the integrand at extended distance, i.e. $> 3L_\kappa$. For the range of $Q_{xy} < Q_{max} = \frac{\pi}{a_m}$, the range of $r'_{xy}$ above which the integrand is no longer oscillating starts around $8L_\kappa$. FIG 11 shows the course of the integrand at two different $Q_{xy}$ values as examples. With the trend of the integrand, it is only necessary to run the integration in $C'(Q_{xy}, \eta, L_\kappa)$ up to around $8L_\kappa$, and makes the numerical integration manageable.

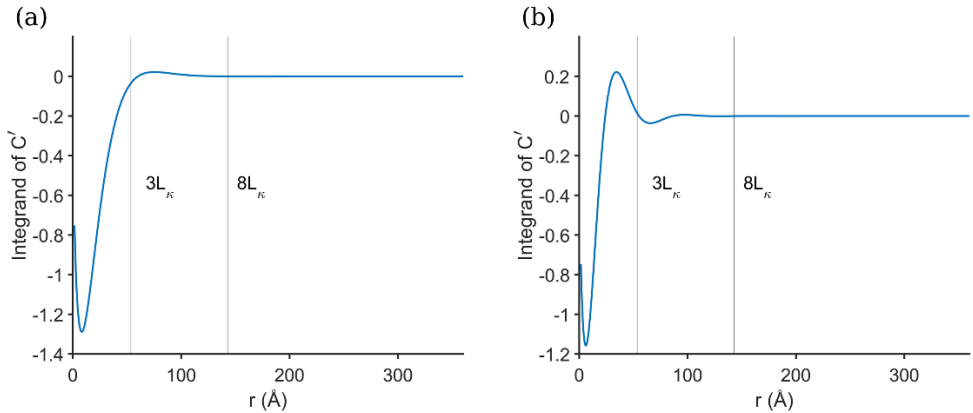

FIG 11. Integrand of $C'$ at $Q_{xy}$ = 0.04 Å$^{-1}$ (a) and 0.1 Å$^{-1}$ (b), both at $Q_z$ = 0.3 Å$^{-1}$. The example sample is set to $\gamma$ = 38 mN/m, $T$ = 293 K, $\kappa$ = 30k$_B$T, and $a_m$ = 5 Å. The two vertical lines mark the distance of $3L_\kappa$ and $8L_\kappa$.

$C(\eta, L_\kappa)$ is the integration of $C'$ around the specular position using circular resolution, and contributes to roughness factor and hence the thermal roughness value. We now evaluate its contribution relative to other terms in the thermal roughness expression (Eq. (23)), to rationalize the rigidity and $Q_z$ range where its contribution can be ignored (Eq. (25)).

*Contact author: chen.shen@desy.de

†Contact author: ocko@bnl.gov

The exponent in the integrand in $C(\eta, L_\kappa)$ increases from 0 asymptotically to unity when $r_{xy} < 8L_\kappa$. The minimum of the integrand decreases with $\kappa$ and approaches roughly on the order of -30 at $\kappa \sim 1000 k_B T$ (a reasonable upper limit for a soft matter thin film). The upper limit of the $|C(\eta, L_\kappa)|$ is thus roughly $10^4$ Å$^{2-\eta}$, that occurs for a very rigid film. The resolution $\delta Q_{xy,R}$ is in the order of some $10^{-4}$ Å$^{-1}$ for reflectivity. Hence for the high rigidity limit of 1000 k$_B$T, the $\frac{\delta Q_{xy,R}^{2-\eta}}{4\pi} C(\eta, L_\kappa)$ term is negligible compared to $\Lambda$ for $\eta < 1$ in the expressions of $\Psi_{R,eCWM}$ and $\sigma_{R,eCWM}$. For commonly studied films with $\kappa < 100$ k$_B$T, the $\frac{\delta Q_{xy,R}^{2-\eta}}{4\pi} C(\eta, L_\kappa)$ term is negligible for the entire range of $\eta < 2$.

### APPENDIX E: Derivation of several small $Q_z$ approximations of the differential roughness factor

Here we show the derivations of two small $Q_z$ approximations of the differential roughness factor. One is the proof that the zero-$Q_z$ limit of the differential scattering cross section must recover the PSD. The other is a phenomenological proof of Eq. (14). Note that there is no strict mathematical proof of Eq. (14) to the best of our knowledge.

The fact that the zero $Q_z$ limit of $\frac{d\sigma}{d\Omega}$ is the PSD is a natural consequence of the Taylor expansion $\exp(Q_z^2 \langle h(\boldsymbol{r}_{xy})h(\boldsymbol{0})\rangle) \approx 1 + Q_z^2 \langle h(\boldsymbol{r}_{xy})h(\boldsymbol{0})\rangle$:

$$\left\{\lim_{Q_z \to 0} \left[\exp(-Q_z^2 \langle h^2 \rangle) \int \exp(Q_z^2 \langle h(\boldsymbol{r}_{xy})h(\boldsymbol{0})\rangle) \exp(i\boldsymbol{Q}_{xy} \cdot \boldsymbol{r}_{xy}) d\boldsymbol{r}_{xy}\right]\right\}\bigg|_{Q_{xy} \neq 0}$$

$$= \exp(-Q_z^2 \langle h^2 \rangle) \int [1 + Q_z^2 \langle h(\boldsymbol{r}_{xy})h(\boldsymbol{0})\rangle] \exp(i\boldsymbol{Q}_{xy} \cdot \boldsymbol{r}_{xy}) d\boldsymbol{r}_{xy}$$

$$= \exp(-Q_z^2 \langle h^2 \rangle) \left[(2\pi)^2 \delta(Q_{xy}) + Q_z^2 \int \langle h(\boldsymbol{r}_{xy})h(\boldsymbol{0})\rangle \cdot \exp(i\boldsymbol{Q}_{xy} \cdot \boldsymbol{r}_{xy}) d\boldsymbol{r}_{xy}\right]$$

$$= \exp(-Q_z^2 \langle h^2 \rangle) \cdot Q_z^2 [A\langle \tilde{h}(\boldsymbol{Q}_{xy})\tilde{h}(-\boldsymbol{Q}_{xy})\rangle]$$

Note the application of the definition that the inverse Fourier transform of $\langle h(\boldsymbol{r}_{xy})h(\boldsymbol{0})\rangle$ is the PSD:

$$\langle \tilde{h}(\boldsymbol{Q}_{xy})\tilde{h}(-\boldsymbol{Q}_{xy})\rangle = \frac{1}{A} \int \langle h(\boldsymbol{r}_{xy})h(\boldsymbol{0})\rangle \cdot \exp(i\boldsymbol{Q}_{xy} \cdot \boldsymbol{r}_{xy}) d\boldsymbol{r}_{xy}$$

Close to zero $Q_z$, applying $\lim_{Q_z \to 0} \exp(-Q_z^2 \langle h^2 \rangle) \approx 1$ yields

$$\left\{\lim_{Q_z \to 0} \left[\exp(-Q_z^2 \langle h^2 \rangle) \int \exp(Q_z^2 \langle h(\boldsymbol{r}_{xy})h(\boldsymbol{0})\rangle) \exp(i\boldsymbol{Q}_{xy} \cdot \boldsymbol{r}_{xy}) d\boldsymbol{r}_{xy}\right]\right\}\bigg|_{Q_{xy} \neq 0} \approx Q_z^2 [A\langle \tilde{h}(\boldsymbol{Q}_{xy})\tilde{h}(-\boldsymbol{Q}_{xy})\rangle]$$

At larger $Q_z$ the differential scattering cross section deviates from the PSD, since the simple Taylor approximation is no longer valid. Eq. (9) must be used to calculate the differential scattering cross section.

Now we will provide the phenomenological proof of Eq. (14). This is an approximation proposed in ref. [26]. It utilizes the Taylor expansion of the differential roughness factor in Eq. (6) for small $Q_z$ to get rid of the complicated Fourier transform and $\langle h(\boldsymbol{r}_{xy})h(\boldsymbol{0})\rangle$, and this yields:

$$\psi(Q_{xy}, Q_z) = \frac{Q_z^2}{16\pi^2 \sin\alpha} \cdot \exp(-Q_z^2 \langle h^2 \rangle) \int \exp(Q_z^2 \langle h(\boldsymbol{r}_{xy})h(\boldsymbol{0})\rangle) \exp(i\boldsymbol{Q}_{xy} \cdot \boldsymbol{r}_{xy}) d\boldsymbol{r}_{xy}$$

$$\approx \frac{Q_z^2}{16\pi^2 \sin\alpha} \cdot \exp\left[-Q_z^2 \int_{Q_{xy}}^{Q_{max}} \frac{A}{(2\pi)^2} \langle \tilde{h}(\boldsymbol{Q}'_{xy})\tilde{h}(-\boldsymbol{Q}'_{xy})\rangle d\boldsymbol{Q}'_{xy}\right] \cdot Q_z^2 [A\langle \tilde{h}(\boldsymbol{Q}_{xy})\tilde{h}(-\boldsymbol{Q}_{xy})\rangle]$$

$$= \frac{Q_z^4}{16\pi^2 \sin\alpha} \cdot \exp\left[-Q_z^2 \int_{Q_{xy}}^{Q_{max}} \frac{A\langle \tilde{h}(\boldsymbol{Q}'_{xy})\tilde{h}(-\boldsymbol{Q}'_{xy})\rangle}{2\pi} Q'_{xy} dQ'_{xy}\right] \cdot [A\langle \tilde{h}(\boldsymbol{Q}_{xy})\tilde{h}(-\boldsymbol{Q}_{xy})\rangle]$$

(A1)

It implies that when applying a Taylor expansion up to the 1$^{st}$ two terms, i.e. $\exp(Q_z^2 \langle h(\boldsymbol{r}_{xy})h(\boldsymbol{0})\rangle) \approx 1 + Q_z^2 \langle h(\boldsymbol{r}_{xy})h(\boldsymbol{0})\rangle$, it is necessary to replace $\langle h^2\rangle$ by a sum of the PSD modes from $Q_{xy}$ to $Q_{max}$ to allow this to be used to a larger $Q_z$ range, e.g. in previous studies [26]. We will use the CWM case to rationalize it and extend it to the eCWM with a finite bending modulus.

The first step of this is to examine the valid $Q_z$ range of the Taylor expansion above, we keep $\exp(-Q_z^2 \langle h^2\rangle)$ for the moment and first examine the CWM surface case. Inserting the CWM's PSD [18] into the equation above gives

$$\left\{\lim_{Q_z \to 0} \left[\exp(-Q_z^2 \langle h^2\rangle) \int \exp(Q_z^2 \langle h(\boldsymbol{r}_{xy})h(\boldsymbol{0})\rangle) \exp(i\boldsymbol{Q}_{xy} \cdot \boldsymbol{r}_{xy}) d\boldsymbol{r}_{xy}\right]\right\}\bigg|_{Q_{xy} \neq 0}$$
$$= \exp(-Q_z^2 \langle h^2\rangle) \cdot Q_z^2 \cdot \frac{k_B T}{\gamma} \frac{1}{L_g^{-2} + Q_{xy}^2}$$

Inserting $\langle h^2\rangle_{CWM}$ (Eq. (4)) gives $\psi_{CWM}$ as:

$$\lim_{Q_z \to 0} \psi_{CWM}\bigg|_{Q_{xy} \neq 0} = \frac{Q_z^4}{16\pi^2 \sin\alpha} \cdot \left(\frac{1}{L_g}\right)^\eta \left(\frac{1}{Q_{max}}\right)^\eta \cdot \frac{k_B T}{\gamma} \frac{1}{L_g^{-2} + Q_{xy}^2} \approx \frac{Q_z^4}{16\pi^2 \sin\alpha} \cdot \left(\frac{Q_g}{Q_{max}}\right)^\eta \cdot \frac{k_B T}{\gamma Q_{xy}^2}$$

Here $L_g^{-2}$ in the denominator is omitted since for typical X-ray measurement $Q_{xy} \gg L_g^{-1}$ [§], and we've replaced $L_g$ at other places by its reciprocal $Q_g$ for convenience. This equation is only valid for $Q_z \approx 0$ ($\eta \approx 0$) and it is not valid for any larger $Q_z$ because it only exhibits $Q_{xy}^{-2}$ dependence instead of $Q_{xy}^{\eta-2}$. It appears clear that the difference between the equation above and the exact $\frac{d\sigma}{d\Omega}$ of the CWM is $Q_g^\eta$ versus $Q_{xy}^\eta$ [31]. This $Q_g$ originates from $\exp(-Q_z^2 \langle h^2\rangle)$ in the first row of Eq. (14). Note that $\langle h^2\rangle$ can be equivalently calculated by summing up all the modes in the PSD [37]:

$$\langle h^2\rangle = \frac{A}{(2\pi)^2} \int_0^{Q_{max}} \langle \tilde{h}(\boldsymbol{Q}_{xy})\tilde{h}(-\boldsymbol{Q}_{xy})\rangle d\boldsymbol{Q}_{xy} = \frac{A}{2\pi} \int_0^{Q_{max}} \langle \tilde{h}(\boldsymbol{Q}_{xy})\tilde{h}(-\boldsymbol{Q}_{xy})\rangle Q_{xy} dQ_{xy} = \frac{k_B T}{2\pi \gamma} \int_0^{Q_{max}} \frac{Q_{xy} dQ_{xy}}{L_g^{-2} + Q_{xy}^2}$$
$$\approx \frac{k_B T}{2\pi \gamma} \ln\left(\frac{Q_{max}}{Q_g}\right)$$

The desired replacement of $Q_g$ by $Q_{xy}$ can be conveniently achieved by replacing the lower boundary of this integral from 0 to $Q_{xy}$:

$$\frac{A}{(2\pi)^2} \int_{Q_{xy}}^{Q_{max}} \langle \tilde{h}(\boldsymbol{Q}_{xy})\tilde{h}(-\boldsymbol{Q}_{xy})\rangle d\boldsymbol{Q}_{xy} = \frac{k_B T}{2\pi \gamma} \ln\left(\frac{Q_{max}}{Q_{xy}}\right)$$

Now, in the first row of Eq. (A1), the $\langle h^2\rangle$ in $\exp(-Q_z^2 \langle h^2\rangle)$ can be replaced by the integral above that starts from $Q_{xy}$. This replacement yields the approximation to the differential scattering cross section of a CWM surface:

*Contact author: chen.shen@desy.de

†Contact author: ocko@bnl.gov

$$\frac{Q_z^2}{16\pi^2 \sin\alpha} \cdot \exp\left[-Q_z^2 \int_{Q_{xy}}^{Q_{max}} \frac{A\langle \tilde{h}(\boldsymbol{Q}'_{xy})\tilde{h}(-\boldsymbol{Q}'_{xy})\rangle_{CWM}}{2\pi} Q'_{xy} dQ'_{xy}\right] \cdot Q_z^2 [A\langle \tilde{h}(\boldsymbol{Q}_{xy})\tilde{h}(-\boldsymbol{Q}_{xy})\rangle_{CWM}]$$

$$= \frac{Q_z^2}{16\pi^2 \sin\alpha} \cdot \exp\left[-Q_z^2 \frac{k_B T}{2\pi\gamma} \ln\left(\frac{Q_{max}}{Q_{xy}}\right)\right] \cdot Q_z^2 \cdot \left(\frac{k_B T}{\gamma} \frac{1}{L_g^{-2} + Q_{xy}^2}\right)$$

$$= \frac{Q_z^4}{16\pi^2 \sin\alpha} \cdot \left(\frac{Q_{xy}}{Q_{max}}\right)^{\eta} \cdot \frac{k_B T}{\gamma} \frac{1}{L_g^{-2} + Q_{xy}^2} \approx \left(\frac{d\sigma}{d\Omega}\right)_{CWM}$$

Since $\psi_{CWM} = \frac{Q_z^2}{16\pi^2 \sin\alpha} \cdot \exp(-Q_z^2\langle h^2\rangle) \int \exp(Q_z^2 \langle h(\boldsymbol{r}_{xy})h(\boldsymbol{0})\rangle_{CWM}) \exp(i\boldsymbol{Q}_{xy} \cdot \boldsymbol{r}_{xy}) d\boldsymbol{r}_{xy}$, we obtain the Eq. (14) for the CWM surface:

$$\exp(-Q_z^2\langle h^2\rangle) \int \exp(Q_z^2 \langle h(\boldsymbol{r}_{xy})h(\boldsymbol{0})\rangle_{CWM}) \exp(i\boldsymbol{Q}_{xy} \cdot \boldsymbol{r}_{xy}) d\boldsymbol{r}_{xy}$$
$$\approx \exp\left[-Q_z^2 \int_{Q_{xy}}^{Q_{max}} \frac{A\langle \tilde{h}(\boldsymbol{Q}'_{xy})\tilde{h}(-\boldsymbol{Q}'_{xy})\rangle_{CWM}}{2\pi} Q'_{xy} dQ'_{xy}\right] \cdot Q_z^2 [A\langle \tilde{h}(\boldsymbol{Q}_{xy})\tilde{h}(-\boldsymbol{Q}_{xy})\rangle_{CWM}]$$

Now we will apply Eq. (14) to derive the limited $Q_z$ approximation form for the eCWM. The integral of the PSD from $Q_{xy}$ is

$$\int_{Q_{xy}}^{Q_{max}} \frac{A\langle \tilde{h}(\boldsymbol{Q}'_{xy})\tilde{h}(-\boldsymbol{Q}'_{xy})\rangle}{2\pi} Q'_{xy} dQ'_{xy} = \frac{k_B T}{2\pi\gamma} \int_{Q_{xy}}^{Q_{max}} \frac{Q'_{xy} dQ'_{xy}}{L_g^{-2} + (Q'_{xy})^2 + L_\kappa^2 (Q'_{xy})^4}$$

$$= \frac{k_B T}{2\pi\gamma}\left[\int_{Q_{xy}}^{Q_{max}} \frac{Q'_{xy} dQ'_{xy}}{L_g^{-2} + (Q'_{xy})^2} - \int_{Q_{xy}}^{Q_{max}} \frac{Q'_{xy} dQ'_{xy}}{L_\kappa^{-2} + (Q'_{xy})^2}\right]$$

$$\approx \frac{k_B T}{2\pi\gamma} \cdot \frac{1}{2}\left[\ln\frac{Q_{max}^2}{Q_{xy}^2} - \ln\frac{Q_{max}^2 + L_\kappa^{-2}}{Q_{xy}^2 + L_\kappa^{-2}}\right] \approx \frac{k_B T}{2\pi\gamma} \ln\frac{1}{Q_{xy} L_\kappa}$$

Here the trick in APPENDIX A is applied to separate the integrand with quartic denominator into two integrands with quadratic denominators. $Q_{max}^2 + L_\kappa^{-2} \approx Q_{max}^2$ and $Q_{xy}^2 + L_\kappa^{-2} \approx L_\kappa^{-2}$ are applied for rigid film and small $Q_{xy}$. Finally the Eq. (15) is obtained:

$$\psi_{eCWM}(Q_{xy}, Q_z) \approx \frac{Q_z^4}{16\pi^2 \sin\alpha} \cdot \exp\left(-\eta \ln\frac{1}{Q_{xy} L_\kappa}\right) \cdot \frac{k_B T}{\Delta\rho_m g + \gamma Q_{xy}^2 + \kappa Q_{xy}^4}$$
$$\approx \frac{Q_z^4}{16\pi^2 \sin\alpha} \cdot \left(\frac{1}{Q_\kappa}\right)^{\eta} \cdot \frac{k_B T}{\gamma} \frac{Q_{xy}^\eta}{Q_{xy}^2 + L_\kappa^2 Q_{xy}^4}$$

Again $L_g^{-2}$ in the denominator is ignored.

*Contact author: chen.shen@desy.de

†Contact author: ocko@bnl.gov

*Contact author: chen.shen@desy.de

†Contact author: ocko@bnl.gov

*Contact author: chen.shen@desy.de

†Contact author: ocko@bnl.gov

*Contact author: chen.shen@desy.de

†Contact author: ocko@bnl.gov